\documentclass[sigconf,nonacm]{acmart}
\AtBeginDocument{%
  }

\newcommand{\term}[1]{{\color{black}#1}}

\newcommand{\Name}{RAM\xspace}

\newcommand{\ignore}[1]{}

\usepackage{hyperref}
\usepackage{xspace}

\usepackage{booktabs}
\usepackage{multirow}
\usepackage{graphicx}
\usepackage{bbding}
\usepackage{subcaption}

\usepackage{multicol}
\usepackage[most]{tcolorbox}
\usepackage{enumitem}
\usepackage{xspace}

\usepackage[table,xcdraw]{xcolor}

\begin{document}

% =========== Response Macro =================
\newcommand{\reviewer}[1]{
\subsection*{Response to Reviewer \#{#1}}
}
\newcommand{\rcomment}[2]{\noindent{\textbf{#1}}: {#2}}
\newtcolorbox{myquote}[1][]{
    colback=black!10,
    colframe=black!10,
    notitle,
    sharp corners,
    % borderline west={1pt}{0pt}{gray!80!black},
    enhanced,
    breakable,
    left=2pt,
    right=2pt,
    top=2pt,
    bottom=2pt,
    ignore nobreak,
}

% \input{reviews}
% \input{response}
% \input{response_discussion}
%%
%% The "title" command has an optional parameter,
%% allowing the author to define a "short title" to be used in page headers.
\title[From Schema to Signal: Retrieval-Augmented Modeling for Relational Data Analytics]{ 
{From Schema to Signal: Retrieval-Augmented Modeling \\for Relational Data Analytics}
}
% Retrieval Augmented Modeling/Learning

% \author{Paper ID: 571}

%%
%% The "author" command and its associated commands are used to define
%% the authors and their affiliations.
%% Of note is the shared affiliation of the first two authors, and the
%% "authornote" and "authornotemark" commands
%% used to denote shared contribution to the research.

\author{Lingze Zeng}
\email{lingze@comp.nus.edu.sg}
\affiliation{%
  \institution{National University of Singapore}
  \city{}
  \state{}
  \country{}
}

\author{Shaofeng Cai}
\email{shaofeng@comp.nus.edu.sg}
\affiliation{%
  \institution{National University of Singapore}
  \city{}
  \state{}
  \country{}
}

\author{Changshuo Liu}
\email{liu717@comp.nus.edu.sg}
\affiliation{%
 \institution{National University of Singapore}
 \city{}
  \state{}
 \country{}
}

\author{Zhongle Xie}
\email{xiezl@zju.edu.cn}
\affiliation{%
  \institution{Zhejiang University}
  \city{}
  \state{}
  \country{}
}

\author{Yuncheng Wu}
\email{wuyuncheng@ruc.edu.cn}
\affiliation{%
  \institution{Renmin University of China}
  \city{}
  \state{}
  \country{}
}

\author{Beng Chin Ooi}
\email{ooibc@zju.edu.cn}
\affiliation{%
  \institution{Zhejiang University}
  \city{}
  \state{}
  \country{}
}
% \author{John Smith}
% \affiliation{%
%   \institution{The Th{\o}rv{\"a}ld Group}
%   \city{Hekla}
%   \country{Iceland}}
% \email{jsmith@affiliation.org}

% \author{Julius P. Kumquat}
% \affiliation{%
%   \institution{The Kumquat Consortium}
%   \city{New York}
%   \country{USA}}
% \email{jpkumquat@consortium.net}

%%
%% By default, the full list of authors will be used in the page
%% headers. Often, this list is too long, and will overlap
%% other information printed in the page headers. This command allows
%% the author to define a more concise list
%% of authors' names for this purpose.
%\renewcommand{\shortauthors}{Trovato et al.}

%%
%% The abstract is a short summary of the work to be presented in the
%% article.
\begin{abstract}

% Relational data stored in RDBMS is foundational to many real-world applications across domains such as e-commerce, finance, and sociality. While deep neural networks (DNNs) have achieved strong performance on tabular data with a single table, extending these models to relational databases is challenging due to the normalized multi-table structure and complex inter-table relationships. Existing approaches often rely strictly on schema-defined graphs, which overlook implicit semantic signals embedded in tuple attributes and suffer from rigid connectivity.

Relational data stored in RDBMS is foundational to applications across domains such as e-commerce, finance, and social media.
While deep neural networks (DNNs) have achieved strong performance on single flattened tables, extending these models to relational databases is challenging due to the normalized multi-table structure and complex inter-table relationships.
Existing approaches often rely strictly on schema-defined graphs, which overlook implicit semantic signals embedded in tuple attributes and suffer from rigid connectivity.

In this work, we propose \textit{Retrieval-Augmented Modeling} (\Name), a new relational data modeling paradigm that learns from both explicit structure and implicit semantics by integrating information retrieval (IR) with graph-based modeling.
RAM treats tuple attributes as terms and uses random walks to construct contextual ``documents'', enabling the use of IR techniques to estimate semantic relevance between tuples and 
% build -> generate
generate a dynamic and data-driven graph.
Building on this, we introduce two retrieval-based augmentation strategies: ATRA, which leverages intra-table relevance for self-supervised contrastive learning to capture fine-grained attribute correlation, and ETRA, which adds inter-table shortcuts between semantically related tuples to enhance graph connectivity and modeling.
This dual-augmentation is further processed by a modular framework with attribute embedding, feature integration, and graph aggregation layers that unifies both attribute and structural information for expressive and flexible representation learning.
Extensive experiments on five real-world relational databases demonstrate that \Name consistently outperforms existing baselines on diverse prediction tasks, establishing a state-of-the-art for relational data analytics.

\end{abstract}

\maketitle

\section{Introduction}
Relational Database Management Systems (RDBMS) have long been the backbone of data storage and management, widely adopted in industries such as e-commerce, finance, and sociality for their efficiency and reliability~\cite{practice,ecommerce,finance,sociality}. 
Relational data in RDBMS is organized in a multi-table format, with a well-defined schema capturing inter-table dependencies through primary-foreign key constraints.
As relational data carries rich information, there is a growing demand for advanced analytics to uncover patterns and extract knowledge within the database, supporting critical applications like forecasting and personalized services~\cite{app1,app2}.
While advanced deep neural networks (DNNs)~\cite{cai2021arm, arik2021tabnet, huang2020tabtransformer} have shown great success in the conventional tabular learning paradigm on single flattened tables, extending them to relational data remains a significant challenge due to the complex multi-table structure.

\begin{figure*}[t] 
\centering 
\includegraphics[width=0.9\textwidth]{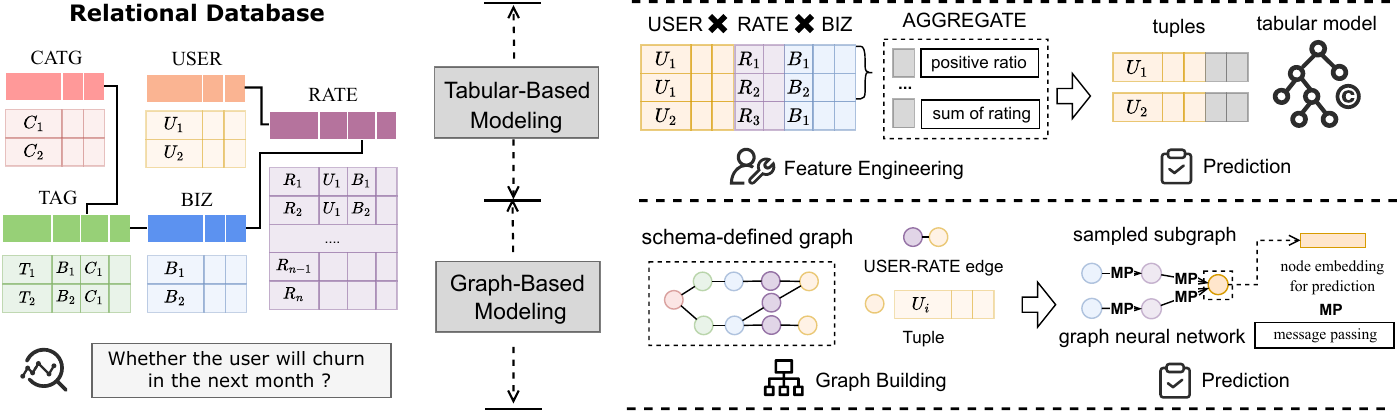}
\caption{Illustration of relational data modeling.}
\label{fig:example}
\end{figure*}

% As illustrated in Figure~\ref{fig:example}, one may forecast whether a user will churn in the next month. The prediction is based on features from the tuple itself and context information from other tables in the relational database.
% %
% In the conventional learning paradigm for tabular data, a single tuple with its attributes is fed into a model for prediction~\cite{shwartz2022tabular,gorishniy2021revisiting,cvetkov2023relational,chepurko2020arda}. 
% Although tabular models can effectively capture the attribute correlations for analytics, applying them to relational data often requires extensive manual preprocessing, 
% such as flattening the multi-table schema around the target table and performing feature engineering to aggregate context information from related tables.
% It compresses relational information into a single, fixed-size representation suitable for tabular models.
% However, as the complexity of inter-table relationships increases, this approach becomes increasingly labor-intensive, error-prone, and leads to suboptimal predictive performance~\cite{lam2017one,lam2018neural,lam2021automated,zhang2023gfs,gan2024graph}.
% sf: [update later] + richer information - flattern tables only contain tuple's own attributes
In a relational database, an analytical task typically aims to predict a specific attribute for a tuple in a target table.
For instance, the task illustrated in Figure~\ref{fig:example} is to predict whether a user will churn based on the user's profile and the interaction history within the database.
The conventional approach for such tasks requires extensive manual preprocessing, most notably, flattening the multi-table schema into a single table centered around the target entity~\cite{shwartz2022tabular,gorishniy2021revisiting,cvetkov2023relational,chepurko2020arda}.
This process compresses relational information into a fixed-size representation applicable for standard tabular models.
However, as the complexity of the database schema increases, the flattening approach becomes labor-intensive, error-prone, and often leads to suboptimal predictive performance due to significant information loss~\cite{lam2017one,lam2018neural,lam2021automated,zhang2023gfs,gan2024graph}.
Specifically, flattening collapses the inherent relational structure by converting fine-grained data from related entities into coarse aggregations, such as reducing a full transaction history to an average purchase amount, while simultaneously discarding rich contextual links defined by both intra-table relationships, e.g., shared preferences among similar users, and inter-table dependencies, e.g., user-product interactions.

Recent advancements in relational analytics have focused on graph-based methods that directly model the complex relationships within a database, thereby circumventing the need for flattening tables~\cite{wang20244dbinfer,relbench}.
In this paradigm, the relational data is transformed into a heterogeneous graph where tuples become nodes, tables define node types, and primary-foreign key relationships form the edges~\cite{zhang2023gfs,cvitkovic2020supervised, atjnet, gan2024graph, pelevska2024transformers, modeling}.
Graph Neural Networks (GNNs) are then applied to this \textit{schema-defined graph} to construct \textit{tuple representations} for predictive analytics by aggregating information from neighboring nodes.
Nonetheless, these GNN-based methods are constrained by their reliance on the explicit schema, leading to two major limitations.
First, the schema-defined graph fails to capture implicit semantic relevance embedded in tuple attributes.
For instance, users who share similar demographic or behavioral attributes may be highly correlated for a prediction task, yet remain disconnected in the graph if no direct PK-FK link exists.
Second, the schema can impose inefficient message-passing routes.
For example, an entity-relationship-entity structure, like USER-RATE-BIZ (user-rating-business) in Figure~\ref{fig:example}, forces information to travel along an indirect multi-hop path via RATE, rather than through a direct one-hop link from USER to BIZ.
The extended routes increase the effective graph diameter, forcing information to travel more hops to cross the graph, which in turn dilutes its content, increases the risk of over-smoothing~\cite{gan2024graph, relbench, oversmoothing}, and impairs the GNN's capacity to model the underlying interaction.

In this paper, we present \textbf{R}etrieval-\textbf{A}ugmented \textbf{M}odeling (\Name), a new relational data modeling paradigm that learns from both explicit structure and implicit semantics in relational data, shifting analytics from schema-bound modeling to a more holistic, dual-representation approach.
Unlike existing graph-based methods that rely entirely on explicit PK-FK links, RAM integrates Information Retrieval (IR) techniques to perceive relationships beyond predefined schema links.
Specifically, RAM constructs a semantic ``document'' for each tuple by combining its attributes with graph context gathered via random walks over the graph.
These documents are then indexed on a per-table basis into an \textit{inverted list} that maps terms to the documents containing them, creating retrieval indices capable of identifying semantically similar documents (tuples).
This enables RAM to dynamically discover and leverage implicit and deep relationships, effectively learning from a data-driven relational graph that complements the static and schema-defined one.
Building on these indices, we introduce two retrieval-based augmentation strategies.
The first, Intr\textbf{A}-\textbf{T}able \textbf{R}etrieval \textbf{A}ugment (ATRA),
% sf: augment -> augmentation, AUGMENT IS A VERB
% sf: can just call it IntraAug, instead of using four letters that are hard to memorize
% sf: same below, InterAug
retrieves semantically similar tuples within the same table, serving as positive pairs for self-supervised contrastive learning.
This encourages the model to learn a more structured and discriminative representation.
The second, Int\textbf{E}r-\textbf{T}able \textbf{R}etrieval \textbf{A}ugment (ETRA), identifies and connects semantically related tuples across tables, even if they are not directly linked by PK-FK constraints.
This enriches the schema-defined graph with new semantic edges, improving graph connectivity and facilitating more efficient message passing between distant yet related entities.

Further, we design a modular, layer-wise modeling framework for relational data modeling.
The framework consists of an \textit{attribute embedding} layer to encode heterogeneous attribute types, a \textit{feature integration} layer to capture intra-tuple feature interactions, and a \textit{graph aggregation} layer to incorporate structural context from the enhanced relational graph.
This layered design effectively decouples the modeling of attribute semantics from that of graph structure, allowing each to be captured independently before unifying them into expressive tuple representations for downstream tasks.

% \begin{figure}[t] 
% \centering 
% \includegraphics[width=0.46\textwidth]{figs/example.pdf}
% \caption{Relational data modeling example.}
% \label{fig:example}
% \end{figure}

We summarize our main contributions as follows.
\begin{itemize}

    \item We present \Name, a novel relational data modeling paradigm that constructs retrieval indices over a database to build a dynamic and data-driven graph for predictive analytics, which learns from both explicit structure and implicit semantics in relational data.

    \item We introduce two RAM strategies: ATRA leverages intra-table semantic similarity for self-supervised contrastive learning to capture fine-grained attribute correlations, while ETRA enriches the graph with inter-table shortcuts to enhance graph connectivity and message passing.

    \item We design a modular GNN-based modeling framework tailored for relational data analytics, which effectively integrates attribute-level semantics with graph structure for expressive tuple representation learning.
    
    \item We conduct extensive experiments on five real-world relational databases across 13 prediction tasks, demonstrating that \Name consistently outperforms 12 baselines, achieving state-of-the-art results.
\end{itemize}

The rest of this paper is structured as follows. We introduce preliminaries in Section~\ref{sec: preliminaries} and detail our retrieval-augmented modeling framework
% sf: be consistent
in Section~\ref{sec: method}, present the experimental evaluation in Section~\ref{sec:exp}, and review related works in Section~\ref{sec:related-work}, followed by the conclusion in Section~\ref{sec:conclusions}.
\section{Preliminaries}
\label{sec: preliminaries}

This section outlines foundational concepts, first defining relational databases and predictive analytics performed on them, then introducing how relational data is modeled as graphs and the basics of information retrieval.

% 1. Relational Data Definition
\subsubsection*{\textbf{Relational Data}}
A relational database is a structured collection of data organized into $K$ tables, $\mathcal{D}:= \{T^k\}_{k=1}^K$.
Each table $T^k$ represents a specific entity type, e.g., customers, products, and contains $N^k$ rows, or \textit{tuples}, and $M^k$ columns, or \textit{attributes}.
We use $T_{i:}^k$ and $T_{:j}^k$ to denote the $i$-th row and $j$-th column in table $T^k$, respectively.

Each row $T_{i:}^k$ represents a single data instance $\mathbf{x_i}$, defined as an ordered set of attribute values $\mathbf{x_i} = (x_1, x_2, \dots, x_{M^k})$.
These attributes can be of heterogeneous types, including numerical, categorical, or textual data.
In contrast to conventional tabular learning, which operates on a single flattened table, relational data involves multiple tables interconnected through \textit{primary-foreign key} (PK-FK) constraints.
A column $T_{:j}^k$ that forms a \textit{primary key} (PK) uniquely identifies each tuple in its table $T^k$, whereas a \textit{foreign key} (FK) column references the primary key of another table $T^{k'}$,
thereby establishing a relationship between tables.

\textit{Predictive analytics} on relational data involves a \textit{target table} $T^k$ and aims to predict a specific attribute for its tuples.
For instance, a customer churn prediction task seeks to predict the \textit{is\_churn} attribute for each tuple in the \textit{Customers} table.
To capture the time-sensitive nature of predictions on dynamic databases, a database snapshot at a given time $t$ is denoted as $\mathcal{D}(t)$.
Formally, the objective is to learn a predictive model $\mathcal{P}$ that maps an input tuple $\mathbf{x}$ from the target table $T^k$ and its database context $\mathcal{D}(t)$ to an outcome $y$:
\begin{align} \label{eq:prediction-proc}
\mathcal{P}(\mathbf{x}, \mathcal{D}(t), T^k) = y,
\end{align}

\noindent
where the model leverages the tuple's own attributes $\mathbf{x}$ and contextual information from related tables within $\mathcal{D}(t)$ to make the prediction.
Our goal is to develop a general modeling framework applicable to a wide range of such analytical tasks over relational data, supporting diverse businesses such as behavior modeling, risk assessment, or transaction prediction.

\subsubsection*{\textbf{Relational Graph Modeling}}
The interconnected structure of a relational database is naturally suited for a graph-based representation.
Modeling the database as a \textit{heterogeneous graph} translates the complex relationships encoded in the schema into explicit structural information.
This enables the application of expressive graph learning techniques to capture the intricate patterns and dependencies within the relational data.

Specifically, each tuple is represented as a \textit{node}, with its source table defining the \textit{node type}.
The primary-foreign key constraints define relationships between tuples, thereby establishing \textit{edges} between nodes, where each unique constraint corresponds to a distinct \textit{edge type}.
For instance, if a tuple $v_i^k$ in table $\mathcal{T}^k$ contains a foreign key referencing the primary key of a tuple $v_j^{k'}$ in table $\mathcal{T}^{k'}$, an \textit{undirected edge}
% sf: directed edge more common, undirected edge also ok
$(v_i^k,v_j^{k'})$ is established.
% sf: these are inter-table edges, what about intra-table edges, namely edges between tuples in the same table?
Formally, this schema-based graph can be defined as:
\begin{align} \label{eq:heterogeneous-graph}
\mathcal{G} = (\mathcal{V}, \mathcal{E}, \mathcal{T_V}, \mathcal{T_E}),
\end{align}

\noindent
where $\mathcal{V} = \cup_{k}\mathcal{V}^k$ is the set of all nodes (tuples), with $\mathcal{V}^k$ being the set of nodes from table $T^k$;
$\mathcal{E} \subseteq \mathcal{V} \times \mathcal{V}$ is the set of edges, defined by PK-FK relationships;
$\mathcal{T_V}$ is the set of node types (tables);
and $\mathcal{T_E}$ is the set of edge types (PF-FK constraints).
% , corresponding to the tables in the database $\mathcal{D}$;
% , corresponding to the different PF-FK constraints.

% Graph-structured data, prevalent in domains like social networks, molecular chemistry, and knowledge graphs, poses challenges for traditional deep learning methods due to its non-Euclidean structure and complex connectivity~\cite{GraphAttribute}. Graph Neural Networks (GNNs)~\cite{inductiveGNN, gat} address this by learning node representations through message passing, where each node aggregates information from its neighbors.

Learning from such graph structures poses challenges for traditional deep learning methods due to their non-Euclidean geometry and complex connectivity~\cite{GraphAttribute}.
\textit{Graph Neural Networks} (GNNs) are a class of models specifically designed for graph learning, which generate a representation for each node by iteratively aggregating information from its local neighborhood.
At each layer $l$, the representation $\mathbf{h}_u^{{l}}$ for a node $u$ is updated based on its own state and messages from its neighbors $\mathcal{N}(u)$:
% \begin{align} \label{eq:representation-update}
% \mathbf{h}_u^{(l+1)} = \operatorname{UPDATE}^{(l)} \left( \mathbf{h}_u^{(l)}, \operatorname{AGGREGATE}^{(l)} \left( \{ \mathbf{h}_v^{(l)} : v \in \mathcal{N}(u) \} \right) \right),
% \end{align}
\begin{align} \label{eq:representation-update}
\mathbf{h}_u^{(l+1)} = \gamma^{(l)} \left( \mathbf{h}_u^{(l)}, \phi^{(l)} \left( \{ \mathbf{h}_v^{(l)} : v \in \mathcal{N}(u) \} \right) \right),
\end{align}

\noindent
where $\phi^{(l)}$ is a differentiable and permutation-invariant \textit{aggregation function}, e.g., mean or sum pooling, that combines neighbor representations $\mathbf{h}_v^{(l)}$, and $\gamma^{(l)}$ is an update function, e.g., a multi-layer perceptron, that merges this aggregated information with the node representation $\mathbf{h}_u^{(l)}$ from the preceding layer.
This message-passing mechanism allows GNNs to effectively integrate contextual signals from across the tables and generate rich, structure-aware representations for each tuple.

\begin{figure*}[t] 
\centering 
\includegraphics[width=\textwidth]{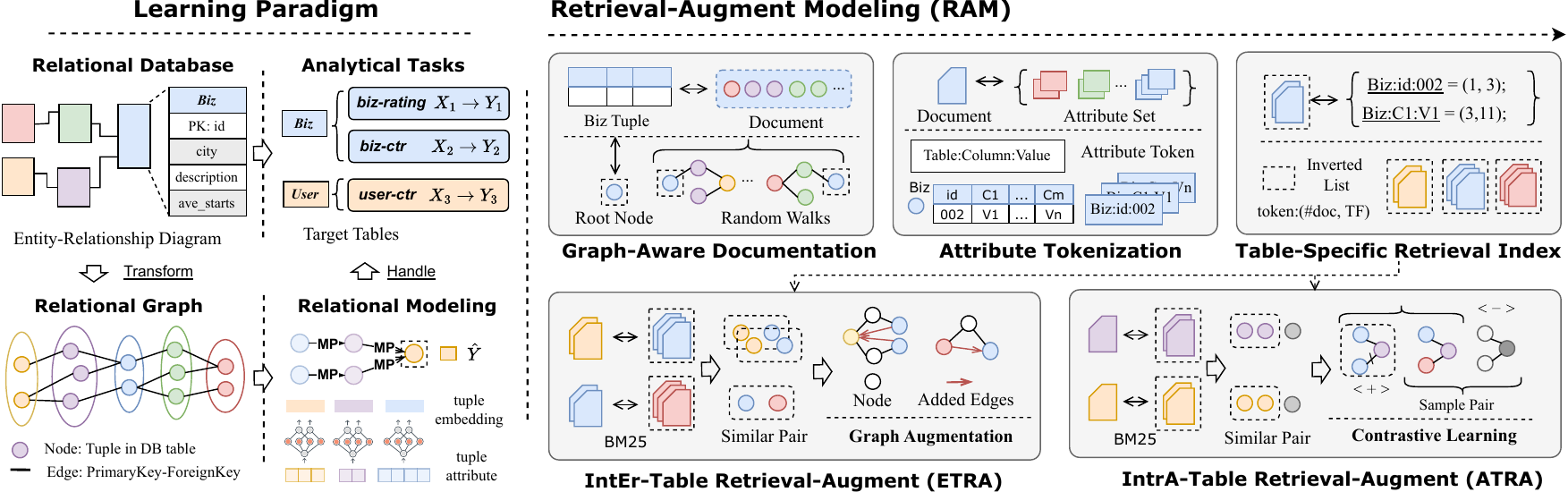} 
\caption{Overview of Retrieval-Augmented Modeling.} 
\label{fig:overview}
\end{figure*}

\subsubsection*{\textbf{Information Retrieval (IR)}}
IR is the task of finding relevant information within large data collections.
Unlike structured database queries requiring exact matches, IR ranks items based on estimated relevance, making them ideal for discovering semantic similarity in applications like search engines and recommender systems.
A foundational approach in IR is the \textit{vector space model}, which often relies on a \textit{Bag-of-Words} (BoW) representation where documents are treated as unordered collections of \textit{terms} (words).
One of the most effective and widely used IR ranking functions is \textit{Best Matching 25} (BM25), a probabilistic model that scores the relevance of a document $D$ to a query $Q$ composed of terms $(q_1, q_2,\cdots, q_n)$:

\begin{align} \label{eq:relevance-score}
    Score(D,Q) \!=\! \sum_{i=1}^{n} \mathbf{IDF}(q_i) \cdot \frac{\mathbf{TF}(q_i, D) \!\cdot\! (k_1 + 1)}{\mathbf{TF}(q_i,D) + k_1 \!\cdot\!(1-b+b \!\cdot\! \frac{|D|}{\rm{avgdl}})},
\end{align}

\noindent
where $\mathbf{TF(q_i, D)}$ is the term frequency of $q_i$ in $D$, $\mathbf{IDF(q_i)}$ is the inverse document frequency of the term (measuring rarity across the collection), $|D|$ is the document length, and $\rm{avgdl}$ is the average document length.
The parameter $k_1$ and $b$ control the term frequency scaling and document length normalization, respectively.

While modern systems increasingly adopt dense retrieval methods based on embeddings from pre-trained language models, BM25 remains a highly efficient and effective technique, particularly for large-scale predictive analytics.
In this work, we adopt these IR principles to estimate semantic relevance between database tuples.
This enables the discovery of implicit signals beyond the rigid database schema, providing a powerful technique for retrieval-based data augmentation.

\section{Retrieval-Augmented Modeling}
\label{sec: method}
In this section, we introduce Retrieval-Augmented Modeling (RAM).
Our approach builds retrieval indices to estimate the relevance between tuples within an RDBMS, extracts informative signals in relational data for augmentation, and supports a neural architecture specifically tailored for relational data analytics.

\subsection{Tuple Retrieval Index}
We describe how to construct a retrieval index for tuples first. As discussed in Section~\ref{sec: preliminaries}, we follow the principles of information retrieval (IR) by treating each tuple as a document, and each attribute as a term.
To support this, we introduce two key components: (1) graph-aware documentation, which constructs a document that captures both the semantic and structural context of a tuple in the relational database; (2) attribute tokenization, which preprocesses attributes to enable meaningful IR statistics, like term frequency. These components are detailed in the following subsections.

\subsubsection*{\textbf{Graph-Aware Documentation}}
We begin by directly taking each tuple as a document, which aligns well with the Bag-of-Word (BoW) assumption in IR, as the attributes are permutation-invariant and agnostic to order and syntax.
However, relying solely on the attributes within a tuple is often insufficient, as it lacks the contextual information present in the relational database. For example, user profile data alone may not provide a complete representation, while incorporating user-product interactions can enhance prediction tasks such as recommendations.
As outlined in Section~\ref{sec: preliminaries}, the database is modeled as a relational graph, where each tuple corresponds to a node. The goal is to aggregate the information from the graph to construct a representative document for the target node (i.e., tuple).
Therefore, we sample relevant nodes around the target node to provide contextual information. Since closer nodes are generally more relevant, they should contribute more to the document. To reflect this, nodes nearer to the target should be sampled more frequently. We employ the Random Walk with Restart (RWR) strategy to generate a root-centric node set, which favors nearby nodes, resulting in higher attribute value counts from closer neighbors.
In RWR, a random walker starts at the root node $r$, and at each step, either moves to a neighboring node or restarts from $r$ with probability $\alpha$. This iterative process converges to a stationary distribution, where each node $v$ is assigned a visitation probability $\pi_v$, indicating its relative relevance to $r$.
Formally:
% \textcolor{red}{
\begin{align} \label{eq:visitation-prob}
\pi_v = \alpha \sum_k^\infty(1-\alpha)^k p_{r\rightarrow v}^{(k)},
\end{align}
% sf: problematic, need to be updated later, e.g., \sum_{k=0}^{\infty}
where $p_{r\rightarrow v}^{(k)}$ is the probability that a length-$k$ pure random walk from $r$ ends at $v$, and $p_{r\rightarrow v}^{(k)} = 0$ for all $k < d(r, v)$. 
% sf: d(r, v) not defined/explained;
% sf: also, in the next sentence, suddenly change d(r, v) to d
This implies $\pi_v \approx \alpha (1-\alpha)^d p_{r\rightarrow v}^{d}$, 
% }
% sf: \alpha (1-\alpha)^{(d)}, (d), to be consistent with the eq.5 (k) notation
indicating that nodes farther from the root receive exponentially lower visitation probabilities. In other words, structurally closer nodes contribute more prominently to the context of the root in RWR sampling.

Given the tuples, we repeatedly apply RWR above the relational graph to generate corresponding documents.
These documents capture both local and multi-hop relational contexts, with node frequency implicitly encoding structural relevance.
As a result, closer nodes contribute more to the representation. 
It facilitates more accurate retrieval and improves the identification of semantically and structurally relevant candidates.

\subsubsection*{\textbf{Attribute Tokenization}}
\begin{figure}
\centering
\includegraphics[width=0.47\textwidth]{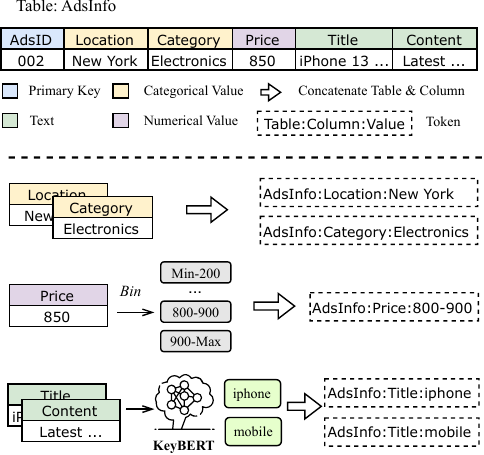}
\caption{Attribute tokenization example.}
\label{fig:tokenize}
\end{figure}

To enable meaningful IR statistics, such as term frequency, attribute values must be discrete. As noted in Section~\ref{sec: preliminaries}, each tuple is represented as a set of heterogeneous attributes, $T^k_{i:} = (x_1, x_2,\cdots,x_{M^k})$. We focus on three common types: numerical value, categorical value, and text. Among them, only the categorical values are inherently discrete and can be directly treated as terms in the BoW model.
Numerical values and text require preprocessing for frequency counting, as illustrated in Figure~\ref{fig:tokenize}.

Specifically, numerical data encompasses continuous values that convey magnitude-based semantics. For example, 8 and 9 are numerically closer than 0 and 9. Treating them as distinct terms ignores this proximity. Directly using raw numerical values as terms may lead to highly sparse representations due to potentially large cardinality. Such sparsity hampers similarity estimation.
Therefore, we adopt a binding strategy
% sf: binding? -> BINNING
to discretize numerical values into a smaller set of intervals, preserving local similarity within each bin.
For simplicity, we apply equidistant binning, where the number of bins is dynamically determined based on the number of distinct values. Following empirical heuristics from prior studies~\cite{binning}, we define the number of bins $b$ as:
\begin{align} \label{eq:bin-number}
b = 
\begin{cases}
1 + \log_2(n), & \text{if } n < 1000 \\
2 \cdot \sqrt[3]{n}. & \text{otherwise}
\end{cases}
\end{align}
For datasets with fewer than 1,000 distinct values, we apply Sturges’ Rule, which assumes a normal distribution and suits smaller datasets. For larger datasets, we use the Rice Rule, which scales better and reduces overfitting.  Once the number of bins is set, the bin boundaries are computed, and numerical values are discretized into terms.

% text
Text snippets, such as user reviews, product descriptions, or comments, are common in relational databases and carry semantic information. 
These unstructured texts are important in various downstream analytical tasks, such as sentiment analysis and user behavior modeling.
However, using raw text as a term directly leads to vocabulary sparsity, since it involves a wide range of word choices and combinations.
Such text values make the similarity score modeling difficult to calculate the term frequency, reducing retrieval quality.
To address this, we apply a topic-based method to condense each text snippet into several representative words that better support term frequency-based retrieval.
In detail, we utilize KeyBERT~\cite{keybert}, a pre-trained topic language model, to identify salient words from the text. These extracted topic words serve as semantically meaningful tokens, reducing vocabulary size while retaining essential semantic content for similarity estimation.

\subsubsection*{\textbf{Table-Specific Retrieval Index}}
Given the graph-aware document and preprocessed attributes, we proceed to construct the retrieval index. A problem arises when identical values appear in different tables or columns, but carry different semantic meanings. For example, in e-commerce, the value ``active'' in the \textit{USER.status} indicates account activity, while in the \textit{BIZ.status} column, ``active'' refers to product availability.
To resolve such ambiguity, we prepend each attribute value with its corresponding table and column name, as shown in Figure~\ref{fig:tokenize}. 
This prefixing strategy ensures that attribute values sharing the same lexical form but originating from different contexts are treated as distinct terms. Consequently, the context of the original schema is preserved, preventing cross-table confusion in similarity estimation.

Subsequently, we build a separate retrieval index for each table, since each table represents a distinct entity. Each index is an inverted list mapping tokens to the tuples (documents) they appear in, along with term frequencies. This facilitates efficient similarity computation using the BM25 ranking function described in Section~\ref{sec: preliminaries}. 
Additionally, given the large number of tuples in relational databases, generating documents for all tables is often impractical. In practice, tables typically fall into two types: entity tables (e.g., users, products) that represent real-world objects, and relationship tables (e.g., interactions, transactions) that capture many-to-many links. Relationship tables are usually larger and less relevant for the prediction task 
which typically target entity-level information.
Therefore, we heuristically classify tables in advance and build retrieval indices only for entity tables, reducing computational overhead while preserving relevance for downstream tasks.

\subsection{Retrieval-Driven Augmentation Signals}
\label{sec:signal}
Based on the constructed table-specific indices, we retrieve tuples from various tables that are semantically and structurally similar to a given query tuple. This enables us to move beyond rigid schema-defined relationships and uncover more flexible, data-driven signals.
In this work, we propose two types of retrieval-driven augmentation.
Intra-Table Retrieval-Augment (ATRA) retrieves relevant tuples within the same table, while Inter-Table Retrieval-Augment (ETRA) identifies related tuples across different tables. These augmentations provide meaningful signals that improve downstream analytical tasks by incorporating contextual patterns not explicitly encoded in the original schema.

\subsubsection*{\textbf{IntrA-Table Retrieval-Augment (ATRA)}}
While the relational graph captures structural dependencies across tables, it often overlooks semantic relationships within a single table.  For instance, users with similar behaviors or products with comparable descriptions may lack explicit links in the schema, yet exhibit strong semantic relevance. These shared properties will lead to similar outcomes in downstream analytical tasks.
To capture such intra-table signals, we use the retrieval index to identify semantically similar tuple pairs within the same table. These pairs are treated as positive samples for self-supervised contrastive learning, encouraging the prediction model to align its representation in the latent space. 

Specifically, for each table with a retrieval index, we sample tuples as queries and retrieve top-ranked, semantically similar tuples from the same table. To identify high-confident positive pairs, we normalize the BM25 scores by dividing each retrieved score by the self-retrieval score of the query tuple, scaling score values between 0 and 1. This normalization facilitates consistent thresholding, and we empirically set the \term{intra-table retrieval threshold} 
% \textcolor{red}{$\theta_a = 0.7$} 
to retain reliable positive pairs. 
% sf: just say \theta_a, report $\theta_a = 0.7$ in experimental setup
Additionally, we apply perturbations to the 
corresponding
% added corresponding
sampled subgraphs during contrastive learning, such as edge removal, attribute masking, and node dropout. The introduced noise makes the model more robust and drives it to learn more resilient and general representations. 
In summary, the augmentation helps the model to capture latent semantic structures and provides a strong initialization for fine-tuning on downstream analytical tasks.

\subsubsection*{\textbf{IntEr-Table Retrieval-Augment (ETRA)}}
The initial graph structure is derived directly from the database schema, where edges represent primary-foreign key constraints. 
% \textcolor{red}{
In many-to-many relationships, this structure requires information to pass through an intermediate relationship table, resulting in two-hop paths between related tuples. These indirect paths weaken the expressive power of the graph and increase the risk of oversmoothing in GNNs, leading to degradation in prediction performance.
% }
% sf: this issue not just occurs in two-hop relationships, but multi-hop relationships -> MORE HOPS -> MORE SEVERE
To address this limitation, we propose an 
% \textcolor{red}{intuitive} 
% sf: intuitive IS WEAK - suggest a lack of rigorous justification, JUST SAY an augmentation strategy
augmentation strategy that identifies highly relevant tuples across different tables
% sf: across different tables -> across tables
and directly links them with new edges. 
Formally, given the relational graph as $\mathcal{G} = (\mathcal{V}, \mathcal{E}, \mathcal{T_V}, \mathcal{T_E})$, the augmented graph is defined as:
\begin{align} \label{eq:augmented-graph}
\mathcal{\hat{G}} = (\mathcal{V}, \mathcal{E} \cup \mathcal{\hat{E}}, \mathcal{T_V}, \mathcal{T_E} \cup \mathcal{\hat{T_E}}),
\end{align}
where $(T^k, T^l) \in \mathcal{\hat{T_E}}$ represents a new edge type between tables $T^k$ and $T^l$, and $D(T^k, T^l) > 1$ 
% sf: again,  $D(T^k, T^l) > 1$ is not defiend here
indicates that they are not directly connected in the original schema. Both $T^k$ and $T^l$ are selected from the set of indexed tables. We focus specifically on table pairs that are indirectly connected in the schema. 

Specifically, for such a table pair, we treat tuples from one table as queries and compute the BM25 function over the index of the other table. 
This yields a similarity score matrix, where top-ranked tuples are considered potential neighbors. We create directed edges from retrieved tuples to the query tuples.
% \textcolor{red}{
% To avoid introducing noisy or weak connections, we apply a dynamic thresholding rule: only retain edges with similarity scores above the \term{inter-table retrieval threshold} $\theta_e = \mu + 2\sigma$ corresponding to approximately 1-95.45\% ~ 4.55\% of edges - 2-sigma percentage, where $\mu$ and $\sigma$ are the mean and standard deviation of the similarity scores, respectively.
% % sf: add this info: corresponding to approximately 1-95.45% ~ 4.55% of edges - 2-sigma percentage
% This ensures that high-confidence links are added. 
% }
To avoid introducing noisy or weak connections, we apply a dynamic thresholding rule that retains only edges whose similarity scores exceed the \term{inter-table retrieval threshold} $\theta_e = \mu + 2\sigma$, where $\mu$ and $\sigma$ denote the mean and standard deviation of the similarity scores, respectively. Under a normality assumption, this threshold retains approximately the top $4.55\%$ of edges, corresponding to the two-sigma tail. This strategy ensures that only high-confidence links are added.
This enhancement shortens effective path lengths, mitigates oversmoothing, and improves the model’s ability to capture cross-table patterns that are not explicitly defined in the original schema.

\begin{figure}
\centering
\includegraphics[width=0.47\textwidth]{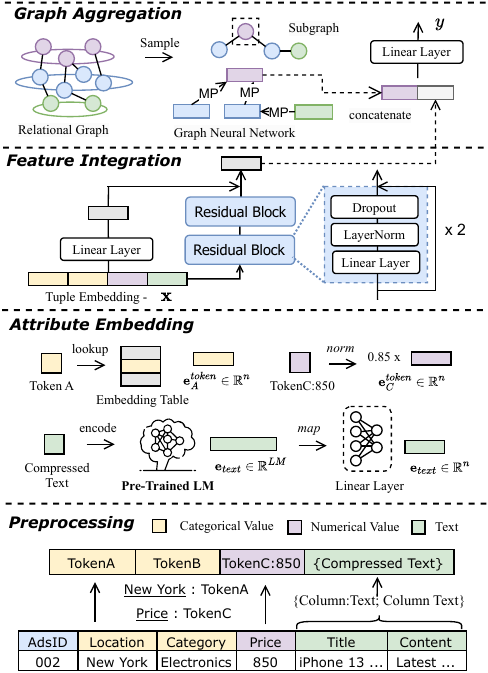}
\caption{Details of relational modeling.}
\label{fig:modeling}
\end{figure}

\subsection{Layer-wise Model Architecture}
In this subsection, we present the architecture of our model for analytical tasks on relational databases. The relational data is first preprocessed and then passes through three layers: the Attribute Embedding layer, which encodes heterogeneous data into a unified latent vector space; the Feature Integration Layer, which captures attribute interactions within tuples to generate tuple-level embeddings; and the Graph Aggregation Layer, which refines these embeddings by incorporating structural information in the database through iterative message passing. Figure~\ref{fig:modeling} illustrates the architecture. We detail the preprocessing step and each layer below, along with the procedure for self-supervised contrastive learning and downstream task fine-tuning.

\subsubsection*{\textbf{Preprocessing}} 
We first encode raw values according to their data types. For categorical and numerical values, we replace each value with a token, with numerical tokens retaining their original values. For text data type, we concatenate all text attributes along with their column names into a single sentence for contextual information. This aggregated sentence replaces the original  text attributes and is embedded as a whole in subsequent stages. Additionally,  primary and foreign key attributes are removed, as they serve only as identifiers while their information is already represented in the attribute values and relational graph structure.
The process is shown at the bottom of Figure~\ref{fig:modeling}.

\subsubsection*{\textbf{Attribute Embedding Layer}}
As described in Section~\ref{sec: preliminaries}, the processed tuple is represented as a value set $\mathbf{x} = (x_1, x_2,\cdots, x_M)$, where each value can be a numerical value, a categorical value, or text. To enable subsequent modeling, any attribute value $x_i$ is transformed into an embedding vector $\mathbf{e}_i$.
We prepare the specific encoding module for each type.
Specifically, categorical values are transformed via \textit{embedding lookup}, i.e., $\mathbf{e}_i = \mathbf{E}_i[x_i]$, $\mathbf{e}_i \in \mathbb{R}^n$, where $n$ is the embedding dimension and $\mathbf{E}_i$ is the embedding table.
Each unique value in $\mathbf{E}_i$ is assigned a distinct embedding vector.
For numerical value $x_j$, embeddings are obtained via a learnable linear transformation, i.e. $\mathbf{e}_j = x_j \cdot \mathbf{\hat{e}}_j + \mathbf{b}_j$, where $\mathbf{\hat{e}_j}, \mathbf{b}_j \in \mathbb{R}^n$, are learnable parameters shared across all values within the same attribute.
% text snippets.
For the aggregated sentence $x_k$, we employ a pretrained language model, which converts $x_k$ into a dense vector representation $\mathbf{e}_k' \in \mathbb{R}^{n_b}$, where $n_b$ is the fixed output dimension of the pre-trained language model. To adapt this representation for downstream modeling, we apply a learnable linear transformation: 
% \textcolor{red}{
$\mathbf{e}_k = \mathbf{W} \cdot \mathbf{e}_k'  + \mathbf{b}$
% }
, where $\mathbf{W} \in \mathbb{R}^{n \times n_b}$, $\mathbf{b}\in\mathbb{R}^n$.
% sf: the formula is informal - the same variable e_k is used to denote both input and output, use e_k' etc to replace one
All data are projected into a unified latent space according to their types.
In this way, a fixed-size representation for each tuple is constructed by concatenating the embeddings of all attributes, i.e. $\mathbf{e} = \mathbf{e}_1 \oplus \mathbf{e}_2 \cdots \oplus \mathbf{e}_M$.

\subsubsection*{\textbf{Feature Integration Layer}}
Given the initial tuple representation $\mathbf{e}$, potential interactions among attributes are not explicitly captured, 
% \textcolor{red}{
despite their importance for accurate tabular predictions. 
% }
% sf: not clear -> e.g., which are crucial for accurate tabular predictions
To capture the interaction among attributes within the tuple, we incorporate a feature integration layer in our model.
Prior studies in tabular learning~\cite{arik2021tabnet,huang2020tabtransformer, cai2021arm} have explored various model designs such as attention or gating strategies. We adopt a simple yet effective approach based on ResNet~\cite{gorishniy2021revisiting}, leveraging residual connections to facilitate stable learning.
Specifically, the tuple representation $\mathbf{e} \in \mathbb{R}^{n_{in}}$ is passed through $n_r$ residual blocks:
\begin{align} \label{eq:residual-blocks}
    \mathbf{e}_f = \underbrace{ResBLK(\cdots ResBLK(\mathbf{e}))}_{n_r}.
\end{align}
Each residual block applies a transformation function $F(\mathbf{x})$ and adds a skip connection with linear projection:
\begin{align} \label{eq:each-block}
ResBLK(\mathbf{x}) = F(\mathbf{x}) + \mathbf{W}\mathbf{x},
\end{align}
where $\mathbf{W} \in \mathbb{R}^{n_{out} \times  n_{in}}$ is a linear projection.
In our implementation, $F$ is a two-layer fully connected network:
\begin{align} \label{eq:two-layer-network}
F(\mathbf{x}) = \delta(\mathbf{W_2} \cdot \delta(\mathbf{W_1}\mathbf{x} + \mathbf{b}_1) + \mathbf{b}_2).
\end{align}
Here, $\mathbf{W}_2 \in \mathbb{R}^{n_{out} \times n_{hid}}$, $\mathbf{W}_2 \in \mathbb{R}^{n_{hid} \times n_{in}}$. $\delta$ denotes a sequence of a nonlinear activation, normalization, and dropout.
The residual mechanism prevents gradient vanishing and enables deeper feature representations.
This layer effectively enhances the tuple representation by capturing feature interactions.
% \textcolor{red}{
We heuristically set $n_r = 2$, $n_{hid} = n_{out} = n$, and apply ReLU activation along with layer normalization.
% }
% sf: again, 1. DOT NOT use 'heuristically' in the paper, e.g., -> empirically; 2. reveal the technical details in experimental setup, instead of the methodology section - else looks ad hoc

\subsubsection*{\textbf{Graph Aggregation Layer}}
After generating tuple embeddings $\mathbf{e}_f$, we introduce the Graph Aggregation Layer to integrate structural information encoded in the relational graph. As introduced in Section~\ref{sec: preliminaries}, the graph is denoted as $\mathcal{G} = (\mathcal{V}, \mathcal{E}, \mathcal{T}_\mathcal{V}, \mathcal{T}_\mathcal{E})$. To accommodate the heterogeneous nature of relational data, we extend the GraphSAGE framework~\cite{inductiveGNN} for multi-relation settings.
Formally, given a node $v$ with its embedding at layer $k$ as $\mathbf{h}_v^k$, the updated embedding at layer $k{+}1$ is computed as:
\begin{align} \label{eq:updated-embedding}
\mathbf{h}_v^{k+1} = \delta ( \mathbf{W}^k \mathbf{h}_v^k +\frac{1}{|\mathcal{T}_\mathcal{E}|} \sum_{r\in \mathcal{T}_\mathcal{E}} \sum_{u \in \mathcal{N}_r(v)} \mathbf{W}_r^k \mathbf{h}_u^k).
\end{align}
Here,
$\mathbf{W}^k \in \mathbb{R}^{n_{k+1}\times n_{k}}$ denotes a learnable linear projection applied to the current node, i.e., a skip connection to preserve its original features.
Meanwhile, $\mathbf{W}_r^k \in \mathbb{R}^{n_{k+1}\times n_{k}}$ represents a relation-specific transformation corresponding to edge type $r$. For simplicity, we set $n_{k+1} = n_k = n$.
The set $\mathcal{N}_r(v)$ comprises the sampled neighbors of node $v$ connected via relation type $r$. 
For aggregation, we sum over neighbors within the same relation to fully capture their influence, then average across different relation types to maintain balanced information flow in heterogeneous graphs.
The tuple representation $\mathbf{e}_f \in \mathbb{R}^n$ is passed through the graph aggregation layer to obtain $\mathbf{e}_g \in \mathbb{R}^n$, which incorporates structural context from the relational graph. The final tuple representation is then formed by concatenating the two vectors: $\mathbf{q} = \mathbf{e}_f \oplus \mathbf{e}_g$, where $ \mathbf{q} \in \mathbb{R}^{2n}$.

\subsubsection*{\textbf{Self-supervised Contrastive Learning}}
Following the above steps, we obtain a comprehensive tuple representation, denoted as $\mathbf{q}_v$ for each node $v$.  To perform self-supervised contrastive learning, we construct positive and negative pairs based on the ATRA introduced in Section~\ref{sec:signal}. If node $v$ has an associated positive pair under ATRA, we randomly sample one such tuple $v^+$, and denote its corresponding representation as $\mathbf{q}^+ = \mathbf{q}_{v^+}$.
If no ATRA is available for node $v$, we generate a synthetic positive by perturbing its local subgraph $\mathcal{N}(v)$. Such perturbations include masking attributes, removing nodes, or deleting edges. The perturbed subgraph $\mathcal{N}(v)^+$ is then used to compute a new representation $\mathbf{q}^+$.
For negative sampling, we randomly select unrelated tuples from the graph and denote their representations as $\mathbf{q}^{-}$. We employ the InfoNCE loss~\cite{infonce} to maximize the similarity between positive pairs while minimizing similarity with negative pairs. The loss function is formulated as:
% \textcolor{red}{
\begin{align} \label{eq:loss-function}
\mathcal{L} = -\log\frac{\exp(\phi(\mathbf{q}, \mathbf{q}^{+})/\tau)}{\sum_{i=0}^N \exp(\phi(\mathbf{q}, \mathbf{q}_i)/\tau)},
\end{align}
% }
% sf: N, q_i not defined
where the function $\phi$ measures similarity, typically cosine similarity, and $\tau$ is a temperature parameter controlling the sharpness of the similarity distribution. This contrastive objective allows the model to learn discriminative tuple representations.

\subsubsection*{\textbf{Downstream Task Fine-tuning}}
For downstream supervised learning tasks, we choose task-specific loss functions. For instance, in binary classification, the objective function is:
% \textcolor{red}{
\begin{align} \label{eq:objective-function}
\mathcal{L}({\hat{y}, y} ) = -\frac{1}{N} \sum_i^N \{ y_i  {\rm log} \sigma(\hat{y_i}) + 
    (1 - y_i) {\rm log} (1 - \sigma(\hat{y_i})) \}.
\end{align}
% }
% sf: i=1 - be consistent with below
For regression tasks, we adopt the L1 loss, which is defined as:
% \textcolor{red}{
\begin{align} \label{eq:reg-loss}
\mathcal{L}({\hat{y}, y}) = \frac{1}{N}\sum_{i=1}^N |\hat{y}_i - y_i|,
\end{align}
% }
% sf: y -> yi
where $\hat{y}$ is the prediction label, $y$ is the ground truth label, $N$ is the number of training tuples, and $\sigma(\cdot)$ is the sigmoid function. The prediction $\hat{y}$ is obtained by applying a prediction head to the tuple embedding: $\hat{y} = HEAD(\mathbf{q})$, where $HEAD$ is typically a single-layer linear projection.
% $\mathbf{W} \in \mathbb{R}^{n \times 1}$.

% sf: relations between ATRA/ETRA with contrastive-learning/fine-tuning/inference should be clearly & separately introduced - readers need to be very familier with these techniques and processes to understand what is the pipeline, how they are used, and why

\section{Experiments} \label{sec:exp}
In this section, we evaluate the effectiveness of \Name, using five real-world datasets across different domains. We first introduce the experimental setup and then report the evaluation results.

\subsection{Experimental Setup}

\subsubsection*{\textbf{Datasets}}
We conduct experiments based on the relbench library~\cite{relbench} on five multi-table datasets with complex relationships, drawn from domains of healthcare, sociology, and e-commerce. The statistics of these datasets are summarized in Table~\ref{tab:dataset}, and the corresponding analytical tasks are illustrated in Table~\ref{tab:tasks}.

\noindent(1) \textbf{Trial}~\cite{trial}, a clinical trial database from the AACT initiative that contains detailed records of medical studies for health and treatment research. It involves one classification task to predict whether a trial will succeed (\textbf{study-outcome}) and two regression tasks to estimate the number of affected patients (\textbf{study-adverse}) and the success rate of the trial site (\textbf{site-success}).

\noindent(2) \textbf{Avito}~\cite{avito}, an online advertisement database capturing interactions between users and product ads. It includes one classification task: predicting whether a user will click on ads (\textbf{user-clicks}) in the next 4 days, and one regression task to estimate the click-through rate of ads (\textbf{ad-ctr}).

\noindent(3) \textbf{Stack}~\cite{stack}, a Stack Exchange database tracking user activity across Q\&A topics. It includes one classification task, predicting whether a user will earn any new badges (\textbf{user-badge}) in the next 3 months, and one regression task to predict the number of votes a post will receive (\textbf{post-votes}).

\noindent(4) \textbf{Event}~\cite{event}, a recommendation database derived from user data on a mobile app, tracking social plans, actions, and event details. It includes two classification tasks: predicting if a user will attend the same event again (\textbf{user-repeat}) or ignore event invitations (\textbf{user-ignore}), and one regression task to estimate how many events a user will respond to in the next 7 days (\textbf{user-attendance}). 

\noindent(5) \textbf{Beer}~\cite{beer} is a database of beer reviews from users and drinking places. We design two classification tasks: predicting if a user will post more than 10 beer reviews (\textbf{user-active}) and if a place will achieve an average rating above 75  (\textbf{place-positive}) in the next season, and a regression task to predict the positive rating ratio of a beer, where ratings above 3.5 are considered positive (\textbf{beer-positive}).

\begin{table}[]
\centering
\caption{Dataset statistics.}
\label{tab:dataset}
\resizebox{0.95\columnwidth}{!}{%
\begin{tabular}{@{}lcccccc@{}}
\toprule[1.5pt]
Dataset & \#Tab & \#Rel & \#Col & \#Feat & \#Tuple    & Domain     \\ \midrule
Trial   & 15      & 15         & 77       & 5,369     & 5,434,924  & Healthcare \\
Avito   & 8       & 26         & 11       & 187,162   & 20,678,117 & E-commerce \\
Stack   & 9       & 14         & 51       & 1,062,015 & 4,897,438  & Sociology  \\
Event   & 5       & 7          & 117      & 1,651     & 326,268    & E-commerce  \\
Beer    & 9       & 12         & 116      & 8623      & 5,388,603  & Sociology  \\ \bottomrule[1.5pt]
\end{tabular}%
}
\end{table}

\subsubsection*{\textbf{Baseline Methods}}
We categorize the baseline methods into three groups. The first group is \textit{tabular-based} methods, which apply advanced tabular learning models to relational data. Specifically, the target table is joined with directly connected tables to form a single flattened table for prediction. This group includes \textbf{CatBoost}\cite{catboost}, \textbf{LightGBM}~\cite{lightgbm}, \textbf{MLP}, \textbf{ResNet}~\cite{gorishniy2021revisiting} and \textbf{FT-Trans}~\cite{gorishniy2021revisiting}.
The second group, \textit{graph-based} approaches, transforms relational data into a heterogeneous graph to enable the application of GNNs. It includes \textbf{Node2Vec}~\cite{node2vec}, \textbf{R-GCN}~\cite{modeling}, \textbf{R-GAT}~\cite{gat} 
\textbf{HGT}~\cite{hgt}. 
In this group, Node2Vec notably differs from the others, as it ignores tuple attributes and learns node embeddings solely from graph structure.
Since GAT~\cite{gat} does not natively support heterogeneous graphs, we extend it by averaging attention weights across different edge types. We refer to this variant as \textbf{R-GAT}. 
Given the large-scale graph in our setting, these methods are implemented with neighborhood sampling to support scalable and inductive representation learning.
The third group, \textit{graph-pretrain-based} methods, utilizes unsupervised learning techniques to capture intrinsic information from the relational graph. These techniques includes \textbf{DGI}~\cite{deepgraphinfomax}, \textbf{GraphCL}~\cite{graphcl}, and \textbf{BGRL}~\cite{bgrl}. In our setting, an R-GCN model is pre-trained using these methods and subsequently fine-tuned on downstream prediction tasks to improve performance.

\subsubsection*{\textbf{Settings}}
For an fair comparison, we standardize the experimental settings across all models.
Specifically, both the feature embedding size and the hidden layer dimension are set to 128. 
All GNN and DNN models are configured with 2 layers.
For classification tasks, we evaluate performance using the AUC-ROC metric, while for regression tasks, we report Mean Absolute Error (MAE). To ensure stable training in regression settings, dropout layers are deactivated. We apply early stopping based on validation performance: if the metric does not improve for $p$ consecutive epochs, training is terminated. This helps avoid overfitting to noise and reduces unnecessary computation.

Regarding training hyperparameters, we use a learning rate in the range of 1e-3 to 1e-4 and a batch size of 512 for all methods and datasets. All experiments are conducted on a server with a Xeon Silver 4114 CPU @ 2.2GHz (10 cores), 256GB of memory, and 8 GeForce RTX 3090 Ti. Model implementations are based on PyTorch 2.1.0, PyTorch Geometric 2.5.3, and RelBench 1.1.0, with CUDA 11.8.

% Please add the following required packages to your document preamble:
% \usepackage{booktabs}
% \usepackage{graphicx}
\begin{table}[]
\centering
\caption{Prediction tasks statistics.}
\label{tab:tasks}
\resizebox{0.95\columnwidth}{!}{%xx
\begin{tabular}{@{}llllc@{}}
\toprule[1.5pt]
Task            & \multicolumn{1}{c}{\begin{tabular}[c]{@{}c@{}}\#Train\\ Instances\end{tabular}} & \multicolumn{1}{c}{\begin{tabular}[c]{@{}c@{}}\#Valid \\ Instances\end{tabular}} & \multicolumn{1}{c}{\begin{tabular}[c]{@{}c@{}}\#Test\\ Instances\end{tabular}} & \begin{tabular}[c]{@{}c@{}}Imbalance\\ Ratio (\%)\end{tabular} \\ \midrule
study-outcome   & 11,994                                                                          & 960                                                                              & 825                                                                            & 63.75                                                          \\
study-adverse   & 43,335                                                                          & 3,596                                                                            & 3,098                                                                          & -                                                              \\
site-success    & 151,407                                                                         & 19,740                                                                           & 22,617                                                                         & -                                                              \\ \midrule
user-clicks     & 59,454                                                                          & 21,183                                                                           & 47,996                                                                         & 3.87                                                           \\
ad-ctr          & 5,100                                                                           & 1,766                                                                            & 1,816                                                                          & -                                                              \\ \midrule
user-badge      & 3,386,276                                                                       & 247,398                                                                          & 255,360                                                                        & 4.81                                                           \\
post-votes      & 2,453,921                                                                       & 15,6216                                                                          & 160,903                                                                        & -                                                              \\ \midrule
user-repeat     & 3,842                                                                           & 268                                                                              & 246                                                                            & 48.98                                                          \\
user-ignore     & 19,239                                                                          & 4,185                                                                            & 3,949                                                                          & 16.87                                                          \\
user-attendance & 19,239                                                                          & 2,013                                                                            & 1,958                                                                          & -                                                              \\ \midrule
user-active     & 16,656                                                                          & 2,794                                                                            & 3,558                                                                          & 53.18                                                          \\
place-positive  & 11,337                                                                          & 4,570                                                                            & 2,869                                                                          & 38.64                                                          \\
beer-positive   & 45,922                                                                          & 12,858                                                                           & 7,218                                                                          & -                                                              \\ \bottomrule[1.5pt]
\end{tabular}
}
\end{table}

\begin{table*}[]
\centering
\caption{Classification prediction results (ROC-AUC, higher is better).}
\label{tab:classification}
\resizebox{0.99\textwidth}{!}{%
\begin{tabular}{@{}clccccccccccccc@{}}
\toprule[1.5pt]
\multirow{3}{*}{Dataset} & \multicolumn{1}{c}{\multirow{3}{*}{Task}} & \multicolumn{13}{c}{Method/Result (ROC-AUC $\uparrow$)}                                                                                                                                                      \\ \cmidrule(l){3-15} 
                         & \multicolumn{1}{c}{}                      & \multicolumn{5}{c}{Tabular-Based}                                             & \multicolumn{4}{c}{Graph-Based}                                     & \multicolumn{4}{c}{Graph-Pretrain-Based}               \\ \cmidrule(l){3-15} 
                         & \multicolumn{1}{c}{}                      & CatBoost & LightGBM & MLP    & ResNet & \multicolumn{1}{c|}{FT-Trans}         & Node2Vec & R-GCN             & R-GAT  & \multicolumn{1}{c|}{HGT}    & DGI    & GraphCL & BGRL              & \Name           \\ \midrule
Trial                    & study-outcome                             & 0.7003   & 0.7009   & 0.7122 & 0.7021 & \multicolumn{1}{c|}{\underline{0.7130}} & 0.5544   & 0.7007            & 0.7035 & \multicolumn{1}{c|}{0.7018} & 0.6946 & 0.7028  & 0.7053            & \textbf{0.7292} \\ \midrule
Avito                    & user-clicks                               & 0.5545   & 0.5360   & 0.5498 & 0.5435 & \multicolumn{1}{c|}{0.5463}           & 0.6080   & \underline{0.6640} & 0.6597 & \multicolumn{1}{c|}{0.6373} & 0.6531 & 0.6628  & 0.6570            & \textbf{0.6743} \\ \midrule
Stack                    & user-badge                                & 0.6386   & 0.6343   & 0.6004 & 0.6140 & \multicolumn{1}{c|}{0.6111}           & 0.5317   & 0.8711            & 0.8288 & \multicolumn{1}{c|}{0.8710} & 0.8717 & 0.8716  & \underline{0.8748} & \textbf{0.8939} \\ \midrule
\multirow{2}{*}{Event}   & user-repeat                               & 0.6904   & 0.6804   & 0.6889 & 0.6657 & \multicolumn{1}{c|}{0.6600}           & 0.6679   & \underline{0.7659} & 0.7186 & \multicolumn{1}{c|}{0.7582} & 0.7422 & 0.7653  & 0.7555            & \textbf{0.8011} \\
                         & user-ignore                               & 0.8024   & 0.7993   & 0.7825 & 0.7733 & \multicolumn{1}{c|}{0.7499}           & 0.8070   & 0.8146            & 0.7993 & \multicolumn{1}{c|}{0.8090} & 0.7999 & 0.8151  & \underline{0.8155} & \textbf{0.8490} \\ \midrule
\multirow{2}{*}{Beer}    & user-active                               & 0.8987   & 0.9147   & 0.8952 & 0.8996 & \multicolumn{1}{c|}{0.9008}           & 0.7996   & 0.9228            & 0.8854 & \multicolumn{1}{c|}{0.9246} & 0.9163 & 0.9231  & \underline{0.9248} & \textbf{0.9457} \\
                         & place-positive                            & 0.8007   & 0.8143   & 0.8093 & 0.8067 & \multicolumn{1}{c|}{0.8120}           & 0.6110   & 0.8140            & 0.7904 & \multicolumn{1}{c|}{0.8195} & 0.8130 & 0.8162  & \underline{0.8253}  & \textbf{0.8560} \\ \bottomrule[1.5pt]
\end{tabular}
}
\end{table*}

\begin{table*}[]
\centering
\caption{Regression prediction results (MAE, lower is better).}
\label{tab:regression}
\resizebox{0.99\textwidth}{!}{%
\begin{tabular}{@{}clccccccccccccc@{}}
\toprule[1.5pt]
\multirow{3}{*}{Dataset} & \multicolumn{1}{c}{\multirow{3}{*}{Task}} & \multicolumn{13}{c}{Method/Result (MAE $\downarrow$)}                                                                                                                                                                                                           \\ \cmidrule(l){3-15} 
                         & \multicolumn{1}{c}{}                      & \multicolumn{5}{c}{Tabular-Based}                                                                             & \multicolumn{4}{c}{Graph-Based}                          & \multicolumn{4}{c}{Graph-Pretrain-Based}                                    \\ \cmidrule(l){3-15} 
                         & \multicolumn{1}{c}{}                      & \multicolumn{1}{l}{CatBoost} & LightGBM & MLP    & ResNet                     & \multicolumn{1}{c|}{FT-Trans} & Node2Vec & R-GCN    & R-GAT    & \multicolumn{1}{c|}{HGT}    & DGI    & GraphCL              & BGRL                 & \Name                \\ \midrule
\multirow{2}{*}{Trial}   & site-success                              & 0.4227                       & 0.4250   & 0.4387 & 0.4267                     & \multicolumn{1}{c|}{0.4279}   & 0.4547   & 0.3999 & 0.3913 & \multicolumn{1}{c|}{0.3899} & 0.4100 & 0.3990               & \underline{0.3863}               & \textbf{0.3481}      \\
                         & study-adverse                             & 51.404                       & 44.011   & 50.595 & 46.709                     & \multicolumn{1}{c|}{51.859}   & 52.843   & 44.890 & 45.820 & \multicolumn{1}{c|}{44.931} & 44.385 & 44.052               & \underline{43.584}               & \textbf{42.673}      \\ \midrule
Avito                    & ad-ctr                                    & 0.0408                       & 0.0410   & 0.0426 & 0.0427                     & \multicolumn{1}{c|}{0.0421}   & 0.0417   & 0.0392 & 0.0392 & \multicolumn{1}{c|}{0.0387} & 0.0387 & 0.0381               & \underline{0.0380}               & \textbf{0.0366}      \\ \midrule
Stack                    & post-votes                                & 0.0676                       & 0.0680   & 0.0684 & 0.0682                     & \multicolumn{1}{c|}{0.0687}   & 0.0685   & 0.0673 & 0.0677 & \multicolumn{1}{c|}{0.0679} & \underline{0.0662} & 0.0669               & 0.0678               & \textbf{0.0642}      \\ \midrule
Event                    & user-attendance                           & 0.2635                       & 0.2640   & 0.2642 & 0.2644                     & \multicolumn{1}{c|}{0.2641}   & 0.2733   & 0.2554 & 0.2533 & \multicolumn{1}{c|}{0.2528} & 0.2538 & \underline{0.2518}               & 0.2530               & \textbf{0.2344}      \\ \midrule
\multicolumn{1}{l}{Beer} & beer-positive                             & 0.1780                       & 0.1790   & 0.2115 & \multicolumn{1}{r}{0.2077} & \multicolumn{1}{c|}{0.2068}   & 0.2447   & 0.1769 & 0.2117 & \multicolumn{1}{c|}{0.1782} &    0.1758    & 0.1732 & \underline{0.1697} & \textbf{0.1573} \\ \bottomrule[1.5pt]
\end{tabular}%
}
\end{table*}

\subsection{Experimental Results and Analysis}

\subsubsection*{\textbf{Main Study}}
We compare our method with three groups of baselines with end-to-end experiments across five datasets covering 13 prediction tasks.
Results for classification and regression tasks are summarized in Table~\ref{tab:classification} and Table~\ref{tab:regression}, respectively.

From a macro perspective, graph-based approaches consistently outperform tabular-based methods. 
This gap is due to the limitation of compressing relational data into a single table, which results in the loss of structural information. Even with advanced feature engineering, much of the relational context in the database remains underutilized.
In contrast, graph-based methods preserve the relational structure, enabling more comprehensive utilization of the data during modeling.
In this conclusion, two specific observations are worth highlighting.
First, tabular-based approaches slightly outperform graph-based methods on the study-outcome task in the Trial dataset. This is likely because the target table contains 28 rich, well-defined attributes, sufficient for accurate prediction. 
When the target table provides ample information, tabular-based models excel at capturing attribute interactions and can achieve strong performance.
Second, within the graph-based methods, Node2Vec performs worse than other models, as it relies solely on graph structure and does not consider node attributes. This limits its effectiveness
when attribute information is crucial, such as in the study-outcome task. However, in datasets like Avito, where graph topology carries strong signals (e.g., user clicks, user visits), Node2Vec outperforms tabular-based methods. This highlights the varying importance of structure versus attributes across different datasets.

Next, we compare standard graph-based approaches with graph-pretrained-based methods. We observe in Table~\ref{tab:classification} that pre-training methods offer no significant advantage in classification tasks. For instance, the performance of R-GCN is comparable to its variants initialized with DGI or GraphCL.
However, for regression tasks in Table~\ref{tab:regression}, the graph-pretrained-based methods generally perform better. This discrepancy can be attributed to two main factors.
First, regression requires learning fine-grained, continuous mappings, which are more sensitive to the quality of feature representations. Pre-training provides a stronger initialization that facilitates better generalization and convergence. Second, regression tasks are more prone to overfitting, especially when target values contain noise. The self-supervised training acts as an implicit regularizer, enhancing the robustness and stability of the representation. 

Finally, our proposed method \Name achieves significant improvements over all baseline methods on both regression and classification tasks, demonstrating the effectiveness of our two retrieval-driven augmentation signals.
In Inter-Table Retrieval Augment (ETRA), new edges are added between semantically related nodes, facilitating GNNs to capture high-level dependencies and cross-entity interactions that standard message passing can not reach due to long-hop limitations. This enriches the graph with meaningful links that are not explicitly present in the original schema. 
In Intra-Table Retrieval Augment (ATRA), contrastive learning is employed to align the embeddings of tuples that exhibit semantic relevance based on retrieval scores.
Unlike existing unsupervised graph pre-training methods that rely on structural perturbations for robustness, ATRA leverages intrinsic semantic similarities as supervised signals, enabling the model to learn more informative representations.
This augmentation guides the predictive model to extract generalizable patterns from relational data and helps alleviate over-smoothing in GNN training, resulting in more robust and discriminative embeddings for downstream tasks.

\subsubsection*{\textbf{Ablation Study}}
To better understand the individual contributions of the two augmentations, ETRA and ATRA, we conduct an ablation study on the Trial and Event datasets. By selectively disabling each augmentation in \Name, we evaluate their impact on end-to-end performance across both classification and regression tasks. We use the model without any augmentations as the baseline. The results are summarized in Table~\ref{tab:ab-classification} and Table~\ref{tab:ab-regression}.

We observe that incorporating either augmentation consistently improves the model performance, confirming the effectiveness of our retrieval-driven augmentations. 
Among these two augmentations, ATRA yields more significant gains than ETRA across most datasets and tasks, suggesting it provides stronger inductive signals for representation learning.
ATRA helps the model capture latent patterns and general knowledge inherent in the data while enhancing robustness through perturbation-based contrastive learning.
This robustness is especially valuable in real-world relational databases, where missing or incomplete data is common. In comparison, ETRA focused on graph augmentation, is more sensitive to schema sparsity or weak connectivity. While beneficial, its impact is less pronounced.
These findings highlight the complementary strengths of both augmentations, with ATRA playing a more prominent role in learning transferable and resilient representations for relational data.
\begin{table}[]
\centering
% \captionsetup{justification=centering}
\caption{Ablation study on classification tasks.}
\label{tab:ab-classification}
\resizebox{0.9\columnwidth}{!}{%
\begin{tabular}{@{}llcccc@{}}
\toprule[1.5pt]
\multirow{3}{*}{\begin{tabular}[c]{@{}c@{}}Dataset/ \\Task\end{tabular}} & \multicolumn{1}{c}{Augments} & \multicolumn{4}{c}{Method/Result (ROC-AUC)} \\ \cmidrule(l){2-6} 
                                                                         & \multicolumn{1}{c}{ETRA}   & \XSolidBrush   &  \XSolidBrush    & \CheckmarkBold    & \CheckmarkBold    \\
                                                                         & \multicolumn{1}{c}{ATRA}   &  \XSolidBrush    &  \CheckmarkBold   & \XSolidBrush   & \CheckmarkBold    \\ \midrule
\multicolumn{1}{c}{Trial}                                                & study-outcome                  & 0.7016    & 0.7189    & 0.7080   & 0.7292   \\ \midrule
\multicolumn{1}{c}{\multirow{2}{*}{Event}}                               & user-repeat                    & 0.7718    & 0.7879    & 0.7820   & 0.8011  \\
\multicolumn{1}{c}{}                                                     & user-ignore                    & 0.8253    & 0.8419    & 0.8250   & 0.8490   \\ \bottomrule[1.5pt]
\end{tabular}%
}
\end{table}

% \begin{figure}[t]
% \centering
% \captionsetup{justification=centering}
% \begin{subfigure}[b]{0.225\textwidth}
%     \centering
%     \captionsetup{justification=centering,margin={0cm,0.5cm}}
%     \includegraphics[width=0.96\columnwidth]{figs/loss/loss-event-user-repeat.pdf}
%     \caption{user-repeat.}
% \end{subfigure}~
% \begin{subfigure}[b]{0.225\textwidth}
%     \centering
%     \captionsetup{justification=centering,margin={0cm,0.5cm}}
%     \includegraphics[width=0.96\columnwidth]{figs/loss/loss-trial-study-outcome.pdf}
%     \caption{study-outcome}
% \end{subfigure}

% \begin{subfigure}[b]{0.225\textwidth}
%     \centering
%     \captionsetup{justification=centering,margin={0cm,0.5cm}}
%     \includegraphics[width=0.96\columnwidth]{figs/loss/loss-event-user-attendance.pdf}
%     \caption{user-attendance}
%     \label{fig:sql}
% \end{subfigure}~
% \begin{subfigure}[b]{0.225\textwidth}
%     \centering
%     \captionsetup{justification=centering,margin={0cm,0.5cm}}
%     \includegraphics[width=0.96\columnwidth]{figs/loss/loss-trial-site-success.pdf}
%     \caption{site-success}
% \end{subfigure}
% \caption{Effect of pre-training on training loss trend.}
% \label{fig:loss_trend}
% \end{figure}

\begin{figure}[t]
\centering
\captionsetup{justification=centering}
\begin{subfigure}[t]{0.23\textwidth}
    \centering
    \includegraphics[width=1\columnwidth]{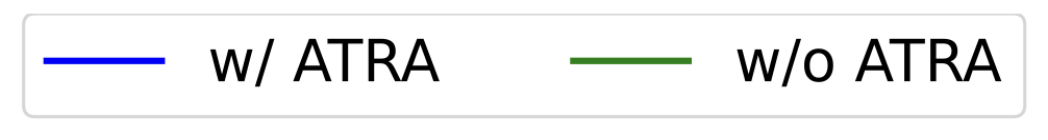}
    % \caption{Classification Tasks}
\end{subfigure}~

\begin{subfigure}[b]{0.235\textwidth}
    \centering
    \includegraphics[width=1\columnwidth]{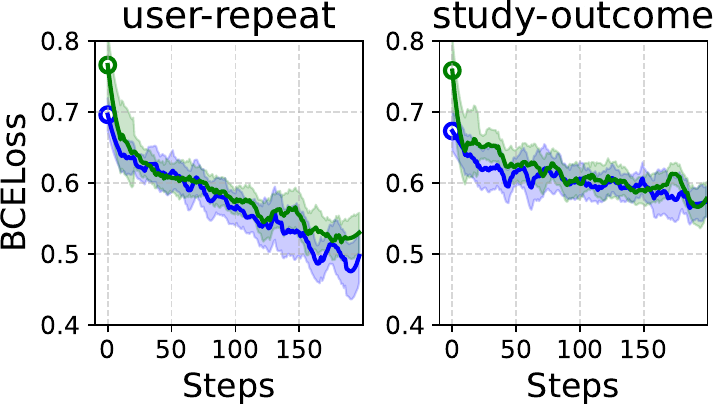}
    \caption{Classification Tasks}
\end{subfigure}
\begin{subfigure}[b]{0.235\textwidth}
    \centering
    \includegraphics[width=1\columnwidth]{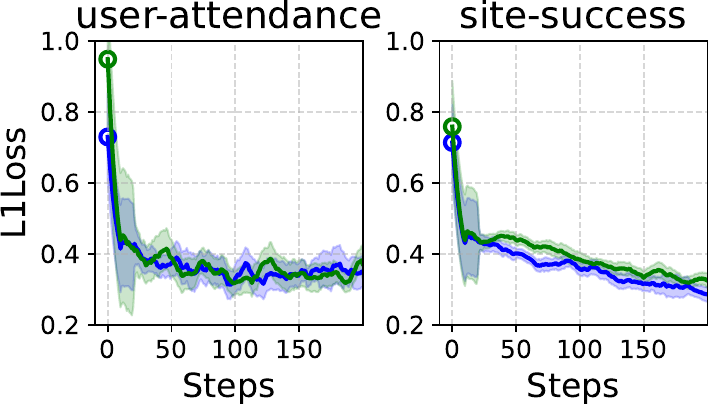}
    \caption{Regression Tasks}
\end{subfigure}
\caption{Effect of ATRA on training loss trend.}
\label{fig:loss_trend_1}
\end{figure}

\begin{figure}[t]
\centering
\captionsetup{justification=centering}
\begin{subfigure}[t]{0.23\textwidth}
    \centering
    \includegraphics[width=1\columnwidth]{figs/legend/pretrain-legend.pdf}
\end{subfigure}~

\begin{subfigure}[b]{0.235\textwidth}
    \centering
    \includegraphics[width=1\columnwidth]{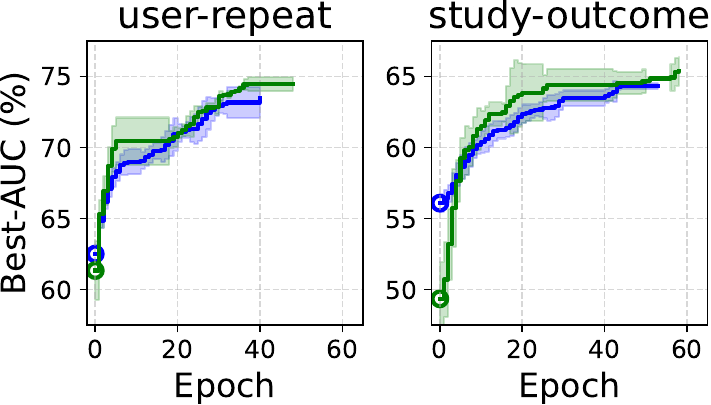}
    \caption{Classification Tasks}
\end{subfigure}
\begin{subfigure}[b]{0.235\textwidth}
    \centering
    \includegraphics[width=1\columnwidth]{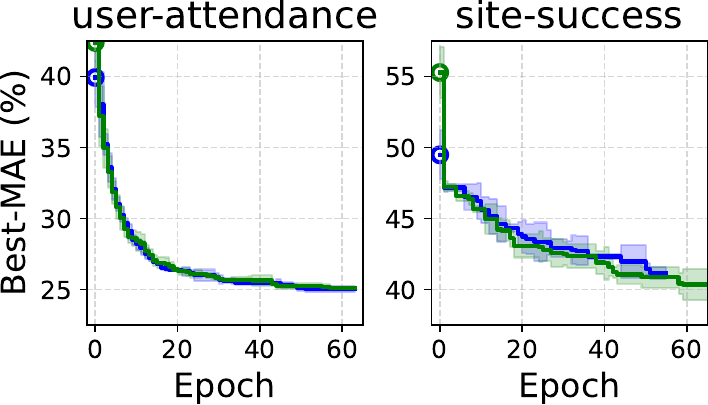}
    \caption{Regression Tasks}
\end{subfigure}
\caption{Effect of ATRA on best validation metrics.}
\label{fig:val_trend}
\end{figure}

Furthermore, we experiment to analyze the effect of ATRA in downstream task fine-tuning. We track the training loss over the first 200 steps, as shown in Figure~\ref{fig:loss_trend_1}, and record the best validation performance achieved in each epoch, presented in Figure~\ref{fig:val_trend}.
In Figure~\ref{fig:loss_trend_1}, we smooth the loss curve and calculate its standard deviation within a sliding window to reduce the impact of stochastic fluctuations and highlight overall optimization trends more clearly. Across datasets and tasks, we consistently observe that the model with ATRA begins with lower initial loss compared to the model without ATRA, indicating that ATRA provides a better initialization and allows the model to start closer to an optimal solution in the parameter space. In Figure~\ref{fig:val_trend}, we further observe that the model with ATRA achieves higher validation performance in early epochs, demonstrating that the initialized representations transfer well and align with downstream objectives. 
As a result of the early-stopping mechanism, the model with ATRA terminated training earlier due to faster convergence and stabilized validation performance. It confirms that self-supervised contrastive learning within ATRA leads to more effective and robust fine-tuning.

\subsubsection*{\textbf{Hyperparameter Study}}
Now, we investigate the effects of three key hyperparameters in \Name: the inter-table retrieval threshold ($\theta_{e}$), which determines the number of newly added edges to the relational graph; the intra-table retrieval threshold ($\theta_{a}$), which filters high-confidence positive pairs for contrastive learning; and the number of negative samples ($N$) used in the contrastive objective.
For $\theta_{e}$, we follow the Chebyshev inequality to define a dynamic threshold based on the mean and standard deviation of retrieved scores. Specifically, we set $\theta_{e} = \mu + k\delta$, where $k$ ranges from 0 to 3, and find that $\mu+2\delta$ yields the optimal performance. 
For $\theta_{a}$, which governs the confidence of positive pairs based on normalized retrieved scores, we explore values from 0.5 to 0.8 and identify 0.7 as the best. We also perform a grid search to determine the optimal number of negative samples $N$ for contrastive learning.

\begin{table}[]
\centering
\caption{Ablation study on regression tasks.}
\label{tab:ab-regression}
\resizebox{0.9\columnwidth}{!}{%
\begin{tabular}{@{}llcccc@{}}
\toprule[1.5pt]
\multirow{3}{*}{\begin{tabular}[c]{@{}c@{}}Dataset/ \\Task\end{tabular}} & \multicolumn{1}{c}{Augments} & \multicolumn{4}{c}{Method/Result (MAE)} \\ \cmidrule(l){2-6} 
                                                                         & \multicolumn{1}{c}{ETRA}   & \XSolidBrush   &  \XSolidBrush    & \CheckmarkBold    & \CheckmarkBold    \\
                                                                         & \multicolumn{1}{c}{ATRA}   &  \XSolidBrush    &  \CheckmarkBold   & \XSolidBrush   & \CheckmarkBold    \\ \midrule
\multicolumn{1}{c}{\multirow{2}{*}{Trial}}                               & site-success                   & 0.3916   & 0.3677  & 0.3775 & 0.3481  \\
\multicolumn{1}{c}{}                                                     & study-adverse                  & 44.137   & 43.024   & 43.651  & 42.673  \\ \midrule
\multicolumn{1}{c}{Event}                                                & user-attendance                & 0.2528   & 0.2474 & 0.2414  & 0.2413 \\ \bottomrule[1.5pt]
\end{tabular}%
}
\end{table}

\begin{figure}[t]
\centering
\begin{subfigure}[b]{0.235\textwidth}
    \centering
    \includegraphics[width=1\columnwidth]{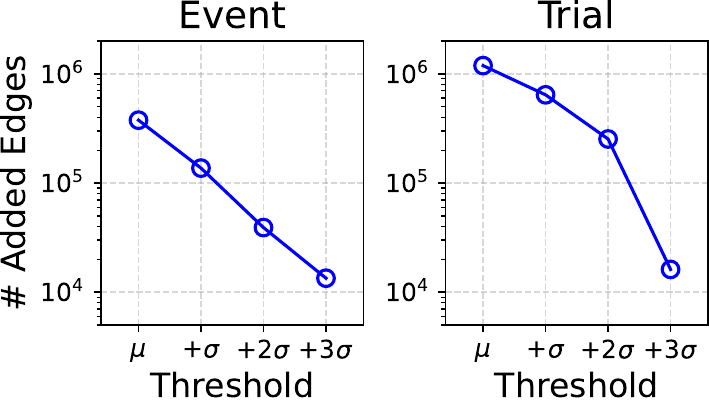}
    \caption{Number of Added Edges}
\end{subfigure}
\begin{subfigure}[b]{0.235\textwidth}
    \centering
    \includegraphics[width=1\columnwidth]{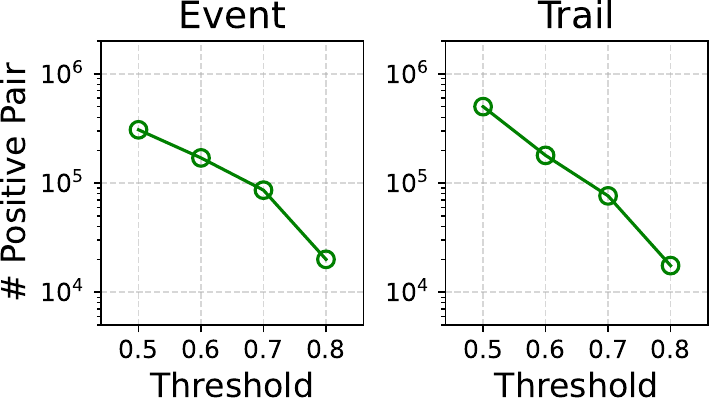}
    \caption{Number of Positive Pairs}
\end{subfigure}
\caption{Effect of $\theta_e$ and $\theta_a$ on the number of signals.}
\label{fig:hyper-trend}
\end{figure}

Figure~\ref{fig:hyper-trend} illustrates how the number of newly added edges and positive pairs varies with changes in $\theta_e$ and $\theta_a$.
As $\theta_e$ increases, more low-similarity pairs are filtered out, preserving only high-confidence candidates. This results in an order-of-magnitude reduction in edge count between $\mu$ and $\mu+3\delta$. 
A similar trend is observed with $\theta_a$. As $\theta_a$ increases, selected positive pairs decrease sharply, reflecting a stricter selection of relevant tuples.

\begin{figure}[t]
\centering
\begin{subfigure}[b]{0.235\textwidth}
    \centering
    \includegraphics[width=1\columnwidth]{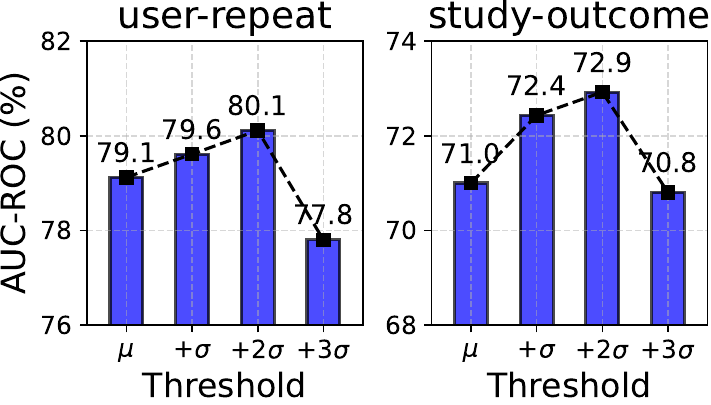}
    \caption{Classification Tasks}
\end{subfigure}
\begin{subfigure}[b]{0.235\textwidth}
    \centering
    \includegraphics[width=1\columnwidth]{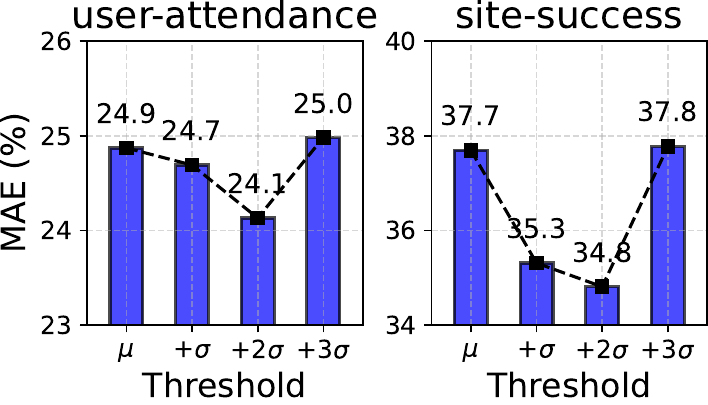}
    \caption{Regression Tasks}
\end{subfigure}
\caption{Effect of threshold $\theta_e$ in ETRA.}
\label{fig:intre-threshold}
\end{figure}

\begin{figure}[t]
\centering
\begin{subfigure}[b]{0.235\textwidth}
    \centering
    \includegraphics[width=1\columnwidth]{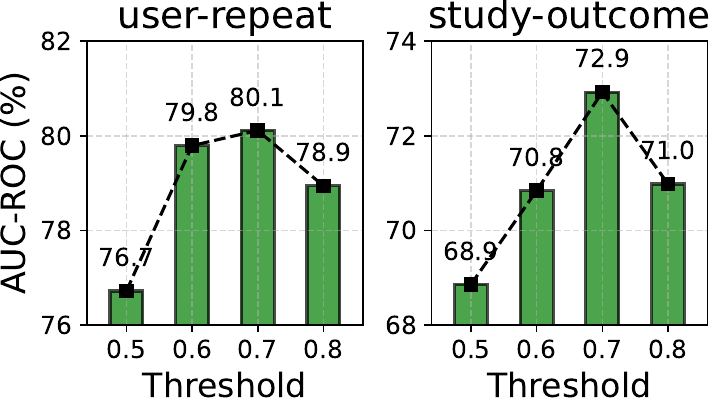}
    \caption{Classification Tasks}
\end{subfigure}
\begin{subfigure}[b]{0.235\textwidth}
    \centering
    \includegraphics[width=1\columnwidth]{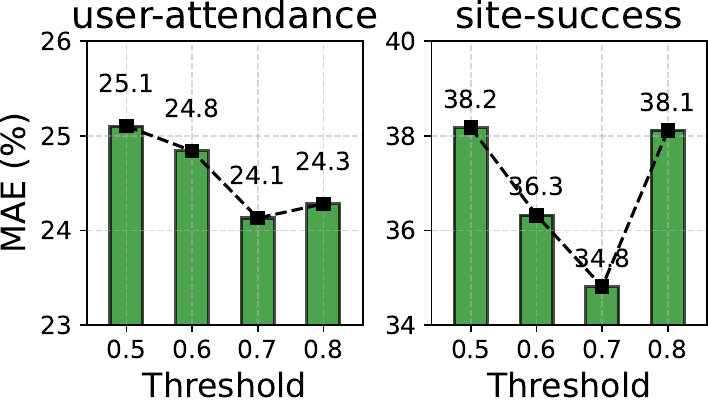}
    \caption{Regression Tasks}
\end{subfigure}
\caption{Effect of threshold $\theta_a$ in ATRA.}
\label{fig:intra_threshold}
\end{figure}

We evaluate the impact of two hyperparameters on model performance using the Trial and Event datasets for both regression and classification tasks. The results are presented in Figure~\ref{fig:intre-threshold} and Figure~\ref{fig:intra_threshold}. We observe a performance improvement as $\theta_e$ increases from $\mu$ to $\mu +2\delta$ and as $\theta_a$ increases from 0.5 to 0.7. It can be attributed to the enhanced signal quality: while the index ranks candidates by similarity scores, not all retrieved pairs are semantically reliable. 
Applying a threshold filters out noisy or weakly related pairs, improving the quality of augmented signals. In contrast, lower thresholds introduce more noise, which can degrade downstream performance.
But, increasing $\theta_e$ beyond $\mu + 2\delta$ or $\theta_a$ beyond 0.7 causes a performance drop due to a sharp reduction in augmented edges and positive pairs. With fewer learning signals, the model benefits less from pretraining and structural learning. These findings highlight a trade-off in threshold selection: higher thresholds improve the signal quality, while overly strict filtering reduces the diversity and quantity of augmentations. Optimal performance lies in balancing signal quality with the richness of augmentation.

\begin{figure}[t]
\centering
\begin{subfigure}[b]{0.235\textwidth}
    \centering
    \includegraphics[width=1\columnwidth]{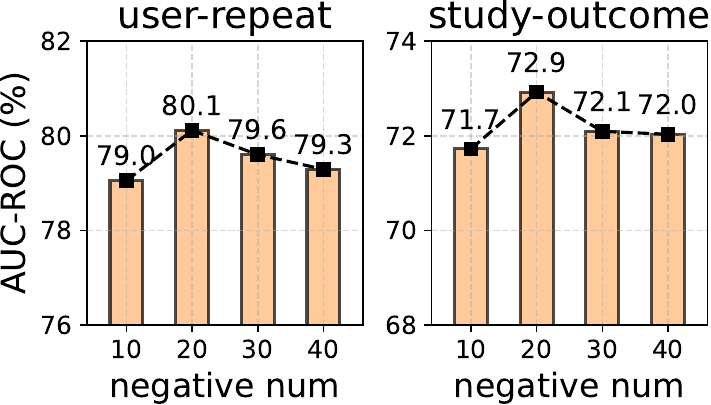}
    \caption{Classification Tasks}
\end{subfigure}
\begin{subfigure}[b]{0.235\textwidth}
    \centering
    \includegraphics[width=1\columnwidth]{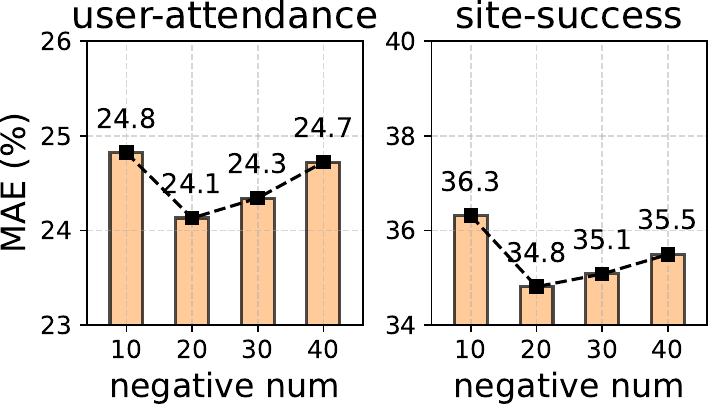}
    \caption{Regression Tasks}
\end{subfigure}
\caption{Effect of the number of negative pairs.}
\label{fig:neg_num}
\end{figure}

We also examine the impact of the number of negative samples $N$ in the contrastive learning process, as shown in Figure~\ref{fig:neg_num}. Unlike the noticeable effect of $\theta_a$ and $\theta_e$, changing $N$ has minimal influence on downstream performance, which remains stable across different values. 
This indicates that a moderate number of negative samples is sufficient for effective contrastive learning, and that adding more negatives contributes little additional benefit. Such robustness means the \Name does not rely heavily on fine-tuning $N$, making it easier to apply across datasets without extensive parameter tuning.

\subsubsection*{\textbf{Interpretability Study}}
To make the effect of \Name more intuitive and straightforward, we directly dive into the effect of ETRA and ATRA using graph-level metrics and visualization methods.

For ETRA, we examine how it changes the structural profile of the relational graph in the Event dataset. 
Specifically, we compute the number of \textit{connected components}, average degree, average shortest path length, and average clustering coefficient before and after applying ETRA. 
These metrics, summarized in Table~\ref{tab:graph-ETRA}, reveal the impact of ETRA on the graph’s topology.
First, we observe a decrease in the number of connected components, suggesting that previously isolated or loosely connected subgraphs are now integrated. This indicates that ETRA enhances the global connectivity of the graph by linking semantically related but distant node pairs. Second, the increase in average degree reflects that nodes have gained additional semantically meaningful neighbors, increasing their local connectivity. Third, the average shortest path length decreases from 14.26 to 12.08, indicating that ETRA introduces semantic shortcut edges, which reduce the number of hops required to traverse the graph. This makes the graph more compact and enables more efficient message passing. Lastly, we observe a substantial rise in the average clustering coefficient from 0.0002 to 0.0631. The initially low value suggests that the original graph had a tree-like or chain-like structure. The sharp increase indicates that the added edges encourage the formation of triadic closures, shifting the graph toward a more locally dense and clustered topology. These new connections foster semantically coherent neighborhoods, further enriching the relational structure.


\begin{figure}
\centering
\includegraphics[width=0.45\textwidth]{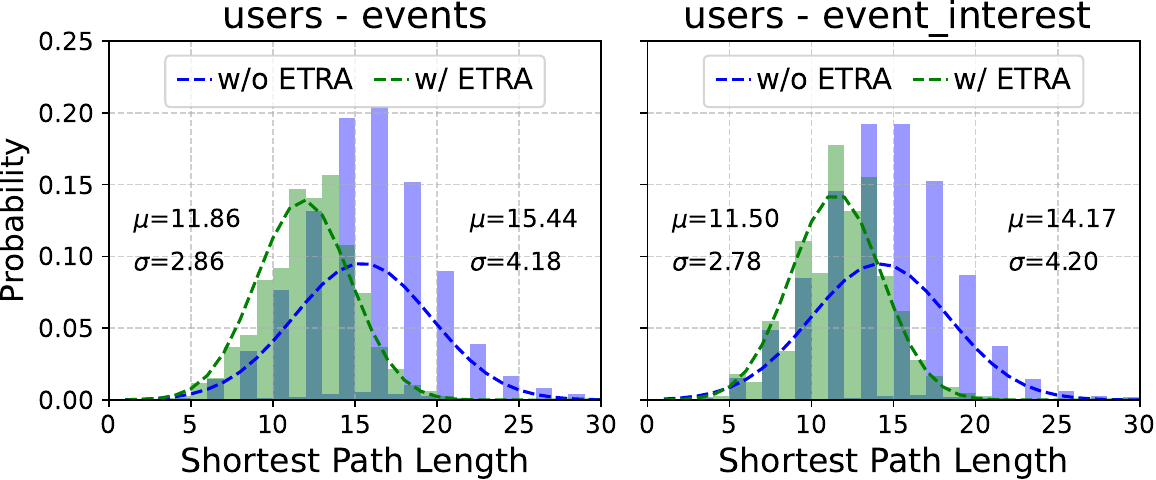}
\caption{Effect of ETRA on the distribution of shortest path lengths for specific paths.}
\label{fig:event_shortest_path_length}
\end{figure}

Additionally, we plot the shortest path length distribution in the Event dataset from the table users to the table events and table event\_interest. To better visualize the change, we fit a normalized distribution to each set of shortest path lengths, shown in Figure~\ref{fig:event_shortest_path_length}. We observe a clear leftward shift in the distribution. For the users-events path, the mean and standard deviation decrease from 15.44/4.18 to 11.86/2.86. Similarly, for users-events\_interest, the values drop from 14.17/4.2 to 11.50/2.78. These shifts indicate a consistent reduction in path lengths and variance, meaning the graph becomes not only more compact but also more uniformly connected, highlighting the effect of ETRA in enhancing semantic connectivity and promoting efficient information flow across the relational graph in GNN.

For ATRA, we list 5 extracted positive pairs from the Tag Table in the Stack Dataset. Due to the instances not going through desensitization, we can evaluate the effect of the ATRA by directly examining the semantic meaning of the tag name. As shown in Table~\ref{tab:positive-pair}, the identified tag groups capture meaningful intra-table relationships. The pair (bic, aic) consists of model selection criteria commonly used in statistical analysis. Set \#2 includes evaluation metrics commonly used in machine learning. The tags in set \#3 pertain to experimental design and modeling approaches in both Bayesian and frequentist frameworks. Set \#4 groups tags related to survival analysis and event-time modeling. Finally, set \#5 includes fundamental concepts and algorithms in reinforcement learning, such as ``Q-learning'' and ``policy-gradient''.
These examples demonstrate that ATRA successfully identifies semantically related tuples. Leveraging these positive pairs in contrastive learning helps improve representation learning by enriching tuple embeddings and enhancing both robustness and diversity.

\begin{figure}
\centering
\includegraphics[width=0.45\textwidth]{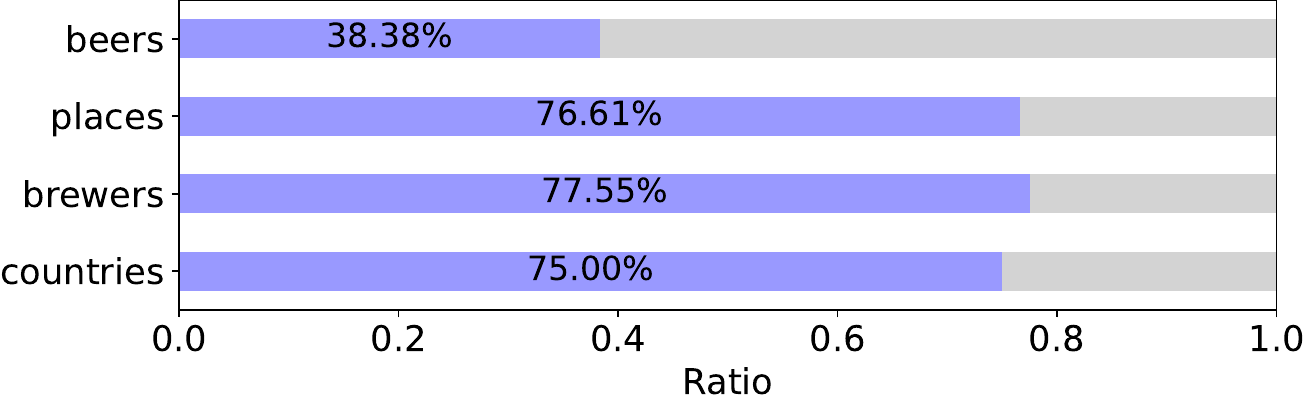}
\caption{
% \textcolor{red}{
Cohort set ratio among ATRA positive pairs in the Beer dataset, with each bar corresponding to a table.
% }
% sf: compare w/ vs. w/o ATRA ratios
}
\label{fig:cohort_ratio}
\end{figure}

To further evaluate the interpretability of ATRA, we conduct an experiment on the Beer dataset to examine whether the extracted positive pairs capture tuples that share implicit common characteristics or belong to the same semantic group, which we refer to as a cohort set~\cite{cohort,cohort2,cohort3}. Specifically, we apply ATRA to four entity tables: \textit{beers}, \textit{places}, \textit{brewers}, and \textit{countries}. We define ground-truth cohort sets based on manually specified patterns. For the \textit{beers} table, tuples are considered to be of the same cohort if they have the same beer style.
For \textit{places}, cohort tuples are those with the same place type or located in the same state. For \textit{brewers}, cohorts share the same brewer type and originate from the same place. For \textit{countries}, cohorts are defined by having the same country code. We compute the cohort set ratio, which measures the proportion of extracted positive pairs that fall within the defined cohort sets. As shown in Figure~\ref{fig:cohort_ratio}, the cohort set ratio exceeds 0.75 for the \textit{places}, \textit{brewers}, and \textit{countries} tables, suggesting that ATRA effectively identifies semantically coherent tuple pairs in these cases. The ratio for the \textit{beers} table is lower, around 0.38, due to the large number (approximately 300) and fine-grained nature of beer styles. The hierarchical relationships among styles make simple style-level matching overly strict, reducing the cohort overlap.

These results demonstrate that ATRA can successfully extract tuple pairs that share implicit semantic similarities
or belong to the same latent cohort. Leveraging these pairs in contrastive learning enables the model to construct a more discriminative latent space for tuple embeddings, thereby enhancing the effectiveness of downstream fine-tuning and prediction tasks.

\section{Related Work} \label{sec:related-work}

\begin{table}[]
\centering
\caption{Impact of ETRA on the graph profile in global view.}
\label{tab:graph-ETRA}
\resizebox{0.75\columnwidth}{!}{%
\begin{tabular}{@{}ccc@{}}
\toprule[1.5pt]
Metrics                      & w/o ETRA & w/ ETRA \\ \midrule
\#Connected Component        & 8584     & 8519    \\
Average Degree               & 3.404    & 3.611   \\
Average Shortest Path Length & 15.436   & 11.861  \\
Average Cluster Coefficient  & 0.0002   & 0.0631  \\ \bottomrule[1.5pt]
\end{tabular}%
}
\end{table}

\subsubsection*{\textbf{Deep Tabular Learning}}
Tabular Learning is designed for data organized in a logical single table, where each instance is a row described by a fixed set of attributes~\cite{shwartz2022tabular,huang2020tabtransformer, arik2021tabnet, gorishniy2021revisiting}.
Classical machine learning approaches such as tree-based models~\cite{lightgbm,catboost} have been widely adopted due to their strong performance and interpretability. Recently, deep learning models for tabular data have emerged, including TabNet~\cite{arik2021tabnet}, FT-Transformer~\cite{gorishniy2021revisiting}, which leverage attention mechanisms or specialized architectures to model complex feature interactions and combinations. 
However, these methods assume that all relevant information is contained within a flat table, making them ill-suited for multiple interconnected tables in a relational database. When cross-table dependencies are complex and essential for prediction, tabular learning struggles to capture the full semantic information. As a result, these methods are only effective when data has already been aggregated or flattened through careful feature engineering.

\subsubsection*{\textbf{Deep Graph Learning}} 
Graph learning methods operate on graph-structured data to capture topological information and dependency patterns. Early approaches, such as PageRank and DeepWalk~\cite{node2vec}, leverage random walks or matrix factorization to model node proximity in the graph. 
Recently, graph neural networks (GNNs) have introduced message-passing mechanisms that allow nodes to aggregate features from their neighbors and learn context-aware embeddings.
Some advanced GNNs, such as  GCN~\cite{gcn}, GAT~\cite{gat}, and GraphSAGE~\cite{inductiveGNN} have been successfully applied to domains like social networks and molecular analysis.
However, GNNs, primarily designed for graph data, do not explicitly model feature interactions and combinations, which limits their ability to uncover the implicit knowledge in tabular data. In relational data, attribute-level semantics are rich and essential. Properly modeling these interactions is critical for accurate predictive analytics.

\begin{table}[]
\centering
% \caption{Examples of similar tag name sets forming positive pairs in ATRA.}
\caption{Example of ATRA positive pairs with Tag table in the Stack dataset.}
\label{tab:positive-pair}
\resizebox{0.75\columnwidth}{!}{%
\begin{tabular}{@{}c|c@{}}
\toprule[1.5pt]
\# & Tag Name Set in ATRA positive pairs\\ \midrule
1 & (bic, aic) \\
2 & (precision-recall, h-measure, auc) \\
3 & (bambi, mixed-model, anova, counterbalancing) \\
4 & (branching, kaplan-meier, survival) \\
5 & (RL, Q-learning, policy-gradient, contextual-bandit) \\ \bottomrule[1.5pt]
\end{tabular}%
}
\end{table}

% \begin{figure}
% \centering
% \includegraphics[width=0.4\textwidth]{figs/visual/tag-links.pdf}
% \caption{Visualization of Positive Pairs from the Tag Table in the Stack Database}
% \label{fig:tag-links}
% \end{figure}

\section{Conclusion} \label{sec:conclusions}

In this work, we propose Retrieval-Augmented Modeling (RAM) for relational data analytics. \Name transforms relational data into a heterogeneous graph and represents each tuple as a document enriched with attributes and graph context. We build information retrieval indices on these documents to estimate semantic relevance between tuples. Leveraging these indices, we introduce two augmentations: Intra-Table Retrieval-Augment (ATRA), which selects similar tuples within tables as positive pairs for contrastive learning, and Inter-Table Retrieval-Augment (ETRA), which links semantically similar tuples across tables to enhance graph connectivity. These augmentations can uncover implicit semantic patterns, thereby addressing the rigid connections inherent in the schema-defined graph. Furthermore, we design a layer-wise model architecture for expressive tuple representation learning. We evaluate RAM on five real-world databases across 13 prediction tasks, where \Name consistently outperforms three groups of baselines, achieving state-of-the-art performance in relational data analytics.

\bibliographystyle{ACM-Reference-Format}
\bibliography{ref}

%%
%% If your work has an appendix, this is the place to put it.
\appendix

% \section{Research Methods}

% \subsection{Part One}

% Lorem ipsum dolor sit amet, consectetur adipiscing elit. Morbi
% malesuada, quam in pulvinar varius, metus nunc fermentum urna, id
% sollicitudin purus odio sit amet enim. Aliquam ullamcorper eu ipsum
% vel mollis. Curabitur quis dictum nisl. Phasellus vel semper risus, et
% lacinia dolor. Integer ultricies commodo sem nec semper.

\end{document}